\documentclass[pdflatex,sn-nature]{sn-jnl}

\usepackage{graphicx}%
\usepackage{multirow}%
\usepackage{amsmath,amssymb,amsfonts}%
\usepackage{amsthm}%
\usepackage{mathrsfs}%
\usepackage[title]{appendix}%
\usepackage{xcolor}%
\usepackage{textcomp}%
\usepackage{manyfoot}%
\usepackage{booktabs}%
\usepackage{algorithm}%
\usepackage{algorithmicx}%
\usepackage{algpseudocode}%
\usepackage{listings}%

\theoremstyle{thmstyleone}%
\theoremstyle{thmstyletwo}%

\theoremstyle{thmstylethree}%

\begin{document}

\title[First principles approaches and concepts for electrochemical systems]{First principles approaches and concepts to simulate electrochemical interfaces}


\author*[1]{\fnm{Mira} \sur{Todorova}}\email{m.todorova@mpie.de}

\author*[1,2]{\fnm{Stefan} \sur{Wippermann}}\email{stefan.wippermann@uni-marburg.de}
\equalcont{These authors contributed equally to this work.}

\author*[1]{\fnm{Jörg} \sur{Neugebauer}}\email{j.neugebauer@mpie.de}
\equalcont{These authors contributed equally to this work.}

\affil*[1]{\orgname{Max Planck Institute for Sustainable Materials}, \orgaddress{\street{Max-Planck-Str. 1}, \city{D\"usseldorf}, \postcode{40237}, \country{Germany}}}

\affil*[2]{\orgname{Philipps-Universit\"at Marburg}, \orgaddress{\street{Department of Physics, \mbox{Renthof 5}}, \city{Marburg}, \postcode{35032}, \country{Germany}}}






\abstract{
\emph{Ab initio} techniques have revolutionised the way in which theory can help practitioners to explore critical mechanisms that govern reactions or properties, and to develop new strategies for materials discovery and design. Yet, their application to electrochemical systems is still limited, due to the challenges electronic structure calculations face in achieving a realistic description of electrified solid/liquid interfaces including, e.g., potential and pH control or free energies of barrier configurations. A well-known example of how novel concepts can extend the scope of simulations is the development of thermostats, which introduced temperature control to electronic structure Density Functional Theory (DFT) calculations. The analogous technique for modelling electrochemical systems – potential control, inherent to most electrochemical experiments – is just emerging.\\

In this review, we critically discuss state-of-the-art approaches to describe electriﬁed interfaces between a solid electrode and a liquid electrolyte in realistic environments. By exchanging energy, electronic charge and ions with their environment, electrochemical interfaces are thermodynamically open systems. In addition, large fluctuations of the electrostatic potential and ﬁeld occur on the time and length scales relevant to chemical reactions. We systematically discuss the key challenges in incorporating these features into realistic \emph{ab initio} simulations, as well as the available techniques and approaches to overcome them, in order to facilitate the development and use of these novel techniques by the wider community. These methodological developments provide researchers with a new level of realism to explore fundamental electrochemical mechanisms and reactions from ﬁrst principles.}




\maketitle
\section{Transferring the macroscopic electrochemical cell into a DFT supercell}\label{sec1}

Achieving an atomic scale understanding of electrochemical processes is imperative to achieve breakthrough innovations. Examples are new generations of supercapacitors \cite{supercap1,supercap2,supercap3,supercap4,supercap5}, metal-air-batteries \cite{mab1,mab2,mab3,mab4,mab5}, transient electronics \cite{transient1,transient2,transient3,transient4}, or new concepts in sustainable metallurgy \cite{plasma,direct,reduction,reduction2}. First principles electronic structure techniques such as DFT are usually the method of choice for studying processes at the atomic scale in condensed systems. While these techniques are routinely employed in many fields, their application to electrochemical systems is still limited, due to challenges that DFT calculations face in achieving a realistic description of electrified solid/liquid interfaces including, for example, potential control, pH control or free energies of barrier configurations.

\begin{figure}[b]%
\centering
\includegraphics[width=0.98\textwidth]{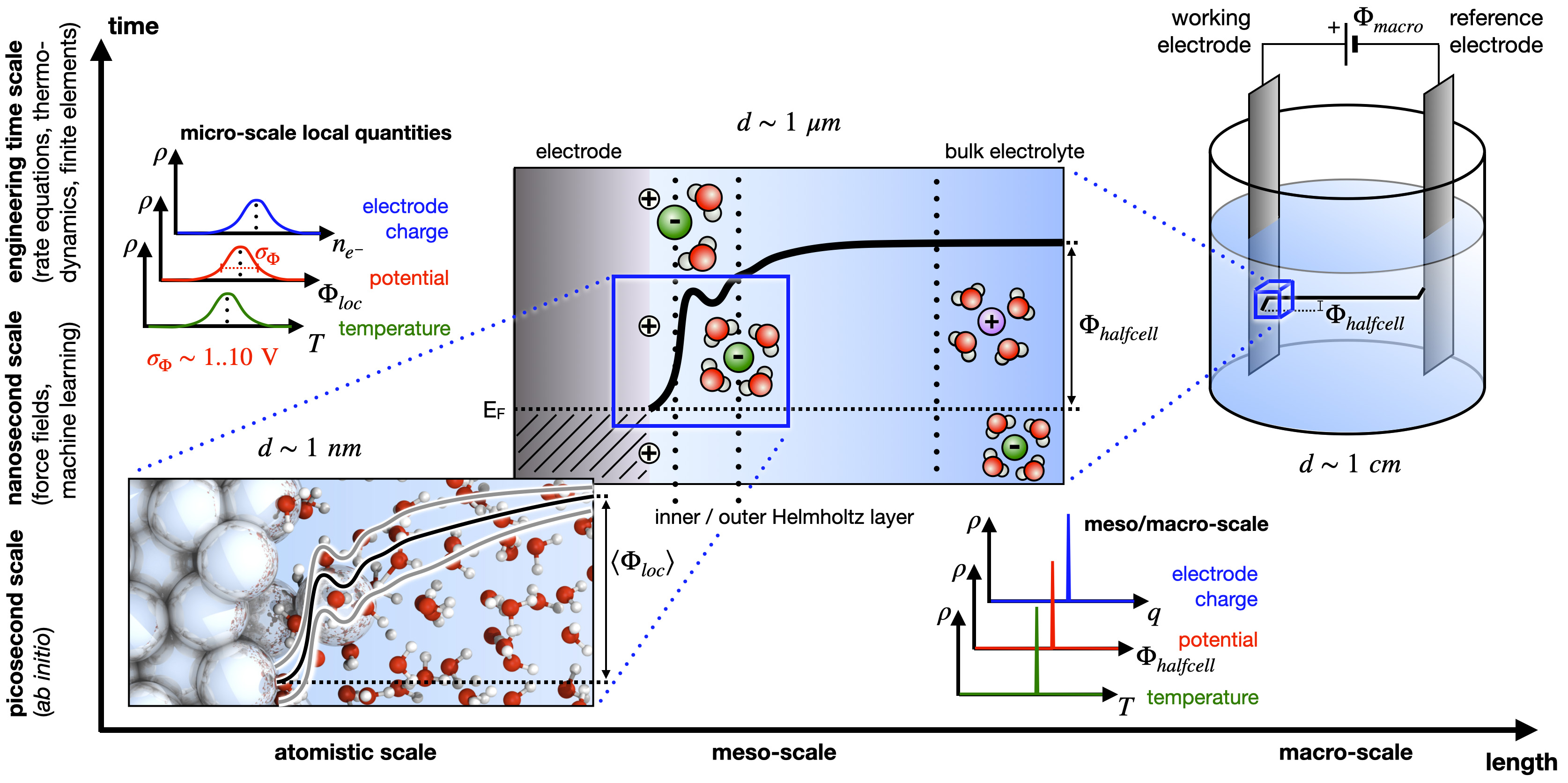}
\caption{Schematic representation of an electrochemical system at the three major length- and time-scales -  macroscopic, mesoscopic and microscopic - together with the corresponding electrostatic potential perpendicular to the solid/liquid interface and laterally averaged. Each of the three electrostatic potentials in the microscopic scale image represents a different configuration obtained from an \emph{ab initio} molecular dynamics simulation. The fluctuation induced distributions of key electrochemical quantities such as surface charge, potential (with $\sigma_\Phi$ being the variance in the potential) and temperature are shown for the microscopic (top-left) and the meso/macro-scale (bottom-right). They show the variation of these quantities over the different snapshots.}\label{fig1}
\end{figure}

A key challenge in describing electrochemical systems and reactions by \emph{ab initio} techniques is the large range of relevant length and time scales \cite{magnussen2019}. The three relevant scales are schematically sketched in Fig.~\ref{fig1}. The actual technical system typically of interest is a full cell consisting of two charged electrodes connected to a potentiostat and immersed in an electrolyte (macroscopic scale). A characteristic feature of electrochemical systems is that the applied voltage in the macroscopic system does not drop homogeneously but only in a small region near the electrode-electrolyte interface, usually referred to as the electric double layer. The induced potential drop near the electrode-electrolyte interface is highly localized, with the strongest electric fields confined to the inner and outer Helmholtz layers, typically within a $\sim$ 1 nm region. This atomistic-scale region is well-suited for atomic scale modelling using modern DFT supercell calculations (cf. Fig. \ref{fig1}, left panel).

However, while DFT can capture the sharp potential drop at the interface, the gradual exponential decay of the potential into the bulk solution - influenced by solvated ions - belongs to the mesoscopic regime (cf. Fig. \ref{fig1}, center panel). We emphasize, however, that any ions which specifically adsorb at the interface or are important in the context of reactions can be straightforwardly included within a DFT supercell calculation. Away from the electrodes the potential is constant, i.e., the system in this region is field free and charge neutral and is -- unlike a capacitor -- not affected by the electrode distance. From a modelling perspective it is thus desirable to capture both the meso-scale decay of the potential outside the DFT supercell and the bulk electrolyte region itself by suitable boundary conditions (cf. also section \ref{sec:boundcon}).

Therefore, it is a key concept to describe the region of the strongest electric fields, crucial for electrochemical and electrocatalytic processes, within DFT while embedding the supercell into its surrounding mesoscopic environment via boundary conditions. To achieve this goal, one must ensure that the bounded finite system described within the DFT supercell replicates the exact dynamics of the unbounded, real physical system. However, due to the microscopic size of the supercell the impact of the exchange of e.g. charge and heat with the environment at the boundaries is large. As a consequence, control parameters like the local potential $\Phi_{loc}$, the surface charge, and the total energy are not constant as for the macroscopic system, where these quantities are lateral averages over a macroscopic area. In the limited lateral dimensions of a typical DFT supercell, these quantities exhibit significant fluctuations due to the small number of particles involved. This behavior is illustrated in Fig.~\ref{fig1} by the density of states, where the distributions broaden compared to the macroscopic case. Each of these control parameters exhibits a normal (Gaussian) distribution due to its thermal origin.

We emphasize that such thermally induced fluctuations in conventional DFT supercells arise solely from their small size. They do not arise from artifacts at lateral boundaries, as bulk water properties can be ensured through performing careful tests of supercell size convergence. We note, however, that local ﬂuctuations at reactive sites remain relevant regardless of the cell size. While their statistical distribution becomes size-independent for sufficiently large cells, the ﬂuctuations themselves do not disappear. In fact, they can still strongly influence the reactivity and require explicit treatment via \emph{ab initio} MD (AIMD) or related methods to be resolved accurately. On the meso- or macroscale, these quantities average over large areas, yielding nearly constant values (delta-like distributions, as shown in the inset). 
Conceptually, these macroscopic averages can be obtained by performing thermodynamically open supercell calculations to capture the fluctuations, followed by averaging over the resulting supercell configurations.

This concept of statistical sampling of physical quantities in the supercell mirrors the use of thermostats in molecular dynamics (MD) simulations \cite{nose1, nose2, berendsen, nhchain, bussi1, bussi2}, which mimic an infinite heat reservoir by allowing energy exchange with the environment to maintain a constant temperature . In electrochemical simulations, the analogous process involves maintaining a stable electrode potential by controlling charge exchange with a surrounding reservoir. In finite-temperature MD, thermostats regulate both the average temperature and the thermodynamically consistent energy fluctuation distribution. This interplay between energy dissipation (temperature control) and fluctuations is fundamentally governed by the fluctuation-dissipation theorem \cite{callen}.

While temperature control with thermostats is common and a well-established practice for MD simulations, the application of potentiostats in 
AIMD simulations is still rare \cite{magnussen2019}. The reason is on one side the difficulty in setting up a suitable surrogate model for the electrochemical double layer. 
On the other side, most computational setups assume constant voltage or constant charge conditions. While these are the appropriate boundary conditions for the mesoscopic/macroscopic case, as shown in Fig.~\ref{fig1}, they cannot capture the sizable voltage and charge fluctuations present on the atomistic level. We note, that the local nature of the fluctuations means that they will have a direct impact on reactions (see discussion in Sec. \ref{sec:boundcon}), which are also locally confined in space and time \cite{amatore2004, santos2024}.

In this review, we start with discussing in Sec. \ref{sec:Electrodes} available concepts to include the impact of the electrochemical double layer in the DFT supercell. These electrostatic surrogate models are used to induce a potential drop across the electrochemical interface between the working electrode and the electrolyte. Sec.~\ref{sec:boundcon} reviews the various electrostatic boundary conditions and their potential shortcomings when applied to the surrogate models discussed in Sec. \ref{sec:Electrodes} within a periodic supercell approach. Sec. \ref{sec:openBC} discusses how to set-up \emph{ab initio} calculations with open thermodynamic boundary conditions, which account for the exchange of electrons with an external reservoir and enable potential control. Sec.~\ref{sec:alignment} focuses on practical challenges in applying electric fields in the context of electronic structure simulations, such as band alignment or dielectric breakdown. Finally, we point in the conclusions to some of the challenges in modelling electrified solid/liquid interfaces, which remain open. 

Further topics are of interest when performing \emph{ab initio} simulations of electrochemical solid/liquid interfaces, but are not covered in this review. We refer readers to the following reviews about e.g. implicit solvents~\cite{rev_reuter} and electrification schemes in implicit solvation~\cite{schwarz2020}, the electric double-layer~\cite{rev_gross}, the use of phase diagrams in modelling corrosion~\cite{rondinelly2019}, electrocatalysis~\cite{koper2012, alfonso2018}, electrochemical energy conversion and storage~\cite{Pastor2024}, energy materials applications~\cite{butler2019} or water at charged interfaces~\cite{rev_gonella}.

\section{Design of a surrogate model for the electrochemical double layer} \label{sec:Electrodes}

\begin{figure}[b]%
\centering
\includegraphics[width=0.98\textwidth]{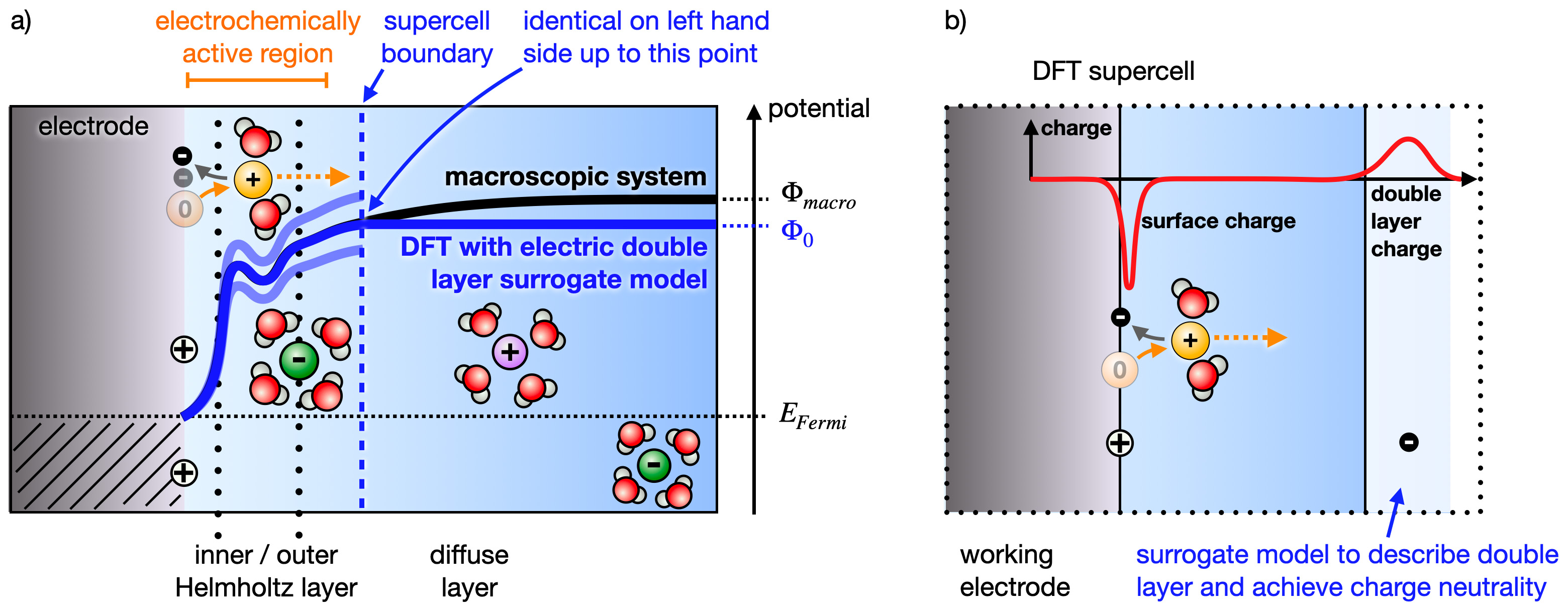}
\caption{\textbf{a)} Comparison between the potential in the macroscopic system and the potential obtained in DFT supercell simulations via a surrogate model for the electric double layer. The potentials need to agree within the microscopic electrochemically active region. To ensure charge neutrality, the residual fraction of the double layer charge that is outside of the supercell must be placed on the double layer surrogate model. Therefore, at larger distances both potentials differ by a constant amount. A suitable correction can be performed \emph{a posteriori}. For the electrostatic potential $\Phi$ we follow the sign convention common to \emph{ab initio} calculations, but note that this is opposite to the sign convention used in classical electrostatics. \textbf{b)} The active region described within the supercell. The potential distribution of the macroscopic system is replicated via introducing a surrogate model for the electrochemical double layer. The surrogate model carries a net charge for compensating the surface charge on the explicit working electrode (see Fig. \ref{fig3} for common realizations of the surrogate model). Note that under reactive conditions, the compensating charge on the surrogate model is not constant but needs to be treated dynamically.}\label{fig6}
\end{figure}

To simulate the electrochemical interface in an \emph{ab initio} supercell, the potential distribution within the simulation cell must replicate that of the real system in this atomistic region, cf. Fig. \ref{fig6}a. In order to achieve this, a charge on the electrode surface or an electric field needs to be applied (Fig.~\ref{fig1}). To maintain charge neutrality under periodic boundary conditions, which is crucial to avoid singularities in the electrostatic energy, the surface charge must be compensated. While many DFT codes address this by introducing a spatially homogeneous background charge, this approach is insufficient for electrochemical systems, where realistic charge compensation must reflect the physics of the electrified interface.

Therefore, in order to achieve realistic field distributions, DFT simulations of electrochemical interfaces commonly employ surrogate models for the electrochemical double layer (cf. Fig. \ref{fig6}b). On the one hand, a primary function of the surrogate model is to compensate for the electrode's surface charge, ensuring charge neutrality of the simulation cell as a whole. On the other hand, beyond merely neutralizing the system, these models also act as boundary conditions, bridging the explicit simulation region with the broader environment of the double layer that lies beyond the DFT supercell. A critical consideration, as outlined in Sec. \ref{sec1}, is that the electric potential difference between the working electrode and the boundary separating the simulation region from the surrogate model cannot remain constant. Due to the thermal motion of ions and molecular dipoles in the double layer, the potential undergoes continuous fluctuations. Consequently, simply matching the average potential at the boundary is insufficient for achieving a realistic description of a highly dynamic and fluctuating system, such as liquid water. Instead, one must also ensure that the distribution of the thermally induced potential fluctuations at the boundary matches that of the (unbounded) physical system.

This concept can be more easily understood by considering an analogy with phonons. For an accurate simulation of heat transport, the boundary of the simulation region must be transparent to phonons, allowing them to pass freely rather than reflecting them back into the system. If the boundary is not transparent, heat would accumulate or reflect, resulting in artificial thermal behavior.

Similarly, for electrochemical simulations, the boundary must be transparent to potential fluctuations. Any reflection or suppression of these fluctuations would alter the dynamics and energetics within the simulation cell, leading to unrealistic field distributions and possibly erroneous predictions for processes such as charge transfer or adsorption. By ensuring transparency to potential fluctuations at the boundary, the surrogate model not only captures the correct mean potential but also replicates the thermal equilibrium behavior of the electric field. This is essential for accurately modeling the electrochemical interface and the dynamic interactions that define it. 

Realizing a suitable surrogate model has been a challenging task. A natural choice would be including an explicit counter electrode as a specific case, in order to mimic the effect of the double layer beyond the supercell. However, most density-functional theory (DFT) simulations for electrode-electrolyte interfaces are performed using supercell calculations in periodic boundary conditions (PBCs)~\cite{che1, filhol2006, taylor2006, schnur, gaigeot, Galli2021, cyli}.
The use of PBCs, which is required in order to describe extended interfaces, strictly enforces (i) simulation cells that are charge neutral in total (due to Gauss's law in electrostatics) and (ii) electric field distributions that obey the periodicity of the supercell. Moreover, (iii) in standard DFT codes there is only a single Fermi level that is constant throughout the supercell. It is thus explicitly impossible to maintain a working electrode and an explicit counter electrode within the same supercell at different Fermi levels in standard DFT simulations.

\begin{figure}[t]%
\centering
\includegraphics[width=0.98\textwidth]{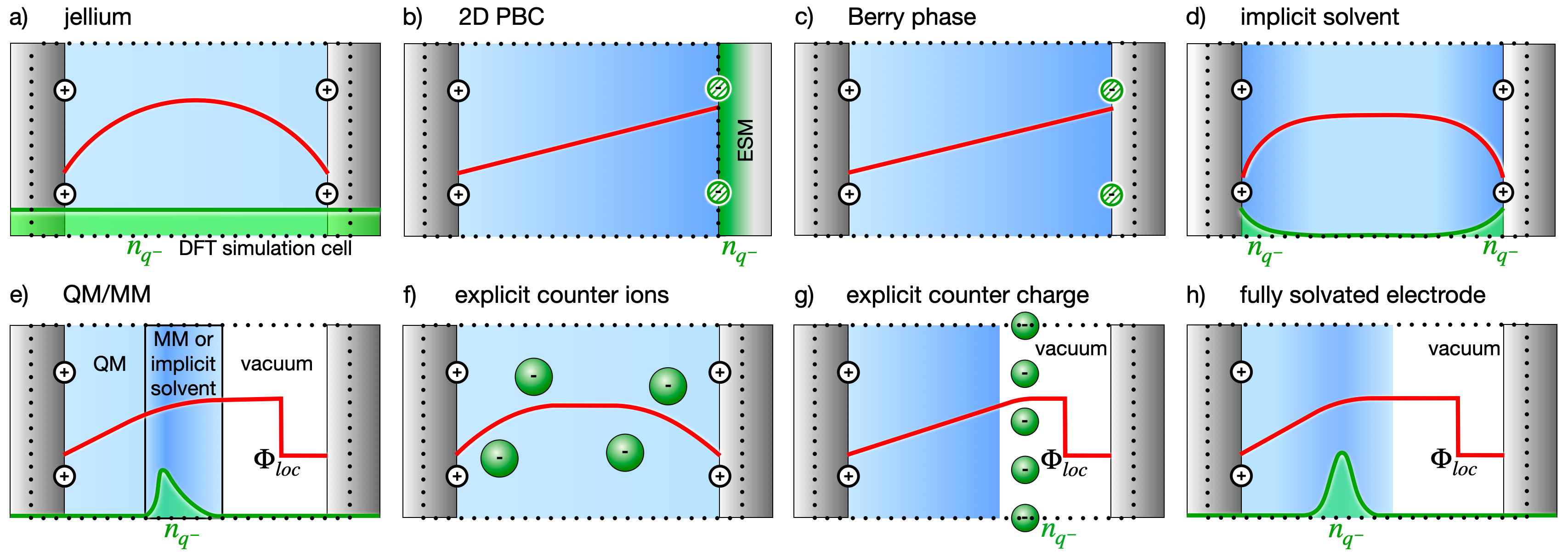}
\caption{The various approaches used to achieve a charged (electrified) electrode-electrolyte interface within a DFT supercell. Electrodes are shown as grey rectangles at the edge(s) of the supercell (marked by a dotted line) and the electrolyte  as a blue area within it. Ions and charges at the electrodes are shown as spheres containing a plus or minus, while atoms are shown as coloured spheres. The resulting electrostatic potential curve is depicted by the red line. The realization and distribution of the countercharge introduced by each of these methods to keep the total charge of the supercell at zero is shown in green (filled spheres represent explicit ions, hatched green spheres represent virtual charges that are a consequence of the respective method and do not need to be added explicitly, and green shapes represent continuous densities, e.g. introduced by a constant background (jellium) in Fig. a).}\label{fig3}
\end{figure}

Therefore, while it is easy in the case of thermostats or barostats to adjust the control parameter (i.e. the kinetic energy or the cell volume, respectively), this is considerably more difficult in the case of potentiostats (electrode charge). Multiple approaches to apply fields or charges in supercells have thus been explored over the past three decades, to address the restrictions outlined above. Fig.~\ref{fig3} provides a schematic overview of the main computational setups.

One of the earliest approaches was suggested by Neugebauer and Scheffler \cite{neugebauer1993}, extending the original dipole correction \cite{neugebauer1992adsorbate} beyond merely compensating for the difference in surface dipoles when using an asymmetric slab. By deliberately introducing a net dipole in the supercell the dipole correction acts like a periodic capacitor -- effectively placing the slab inside a charged capacitor. Since the dipole correction has been implemented in most plane-wave-based DFT codes it provides an easy approach to apply an external potential.
Later, Lozovoi and Alavi \cite{lozovoi} used a symmetric slab setup, cf. Fig.~\ref{fig3}a, where the surfaces are charged by introducing a net charge to the system. In order to ensure charge neutrality of the super cell, an equal and opposite homogeneous background charge is used as a computational counter electrode. While this setup can be easily applied in any DFT code, it does not readily provide a direct measure of the electrode potential (as, for example, the dipole correction \cite{neugebauer1992adsorbate}): in principle, the electrode potential is given by the derivative $\Phi = \frac{\partial \mathcal H}{\partial q}$ of the DFT total energy $\mathcal H$ with respect to the charge $q$, which is, however, a notoriously difficult quantity to compute in 3D PBC supercell calculations \cite{defect_review, FNV_PRL}. 
Also, the distribution of the counter-charge as homogeneous background is very different from the distribution of the counter-charges in a realistic electrochemical cell. In the latter, the counter-charges are realized by the double layer, which shows a distribution very different from a spatially homogeneous and constant background.

Otani \emph{et al.} therefore proposed to use 2D PBCs to describe the lateral periodicity of the interface, while lifting the PBC parallel to the surface normal via Green's function techniques \cite{otani2006esm,greens,otani2013esm2}, cf. Fig.~\ref{fig3}b. Alternatively, one may lift the PBCs selectively only for the electric field in the supercell via electric enthalpy-based approaches \cite{pasquarello2002} or Berry phase techniques \cite{resta} and the modern theory of polarization \cite{stengel2007,stengel2009}, cf. Fig.~\ref{fig3}c. Here, the electrostatic potential drop across the supercell is described by a geometric phase, which accumulates when moving from one periodic image of the supercell to the next. Such approaches to describe electric fields in supercells are demanding to implement. Hence, they are available in only some of the $\sim$ 40+ publicly available DFT codes \cite{sandia}. Moreover, the practical implementations of this approach typically only work for systems with a bandgap. This is due to the challenges in describing fractional occupancies or band crossings using Wannier functions \cite{stengel2007, marzari1997}. Thus, only systems with semiconductor or insulator electrodes can be conveniently studied.

A continuous variation of the charge on the working electrode is achieved via charged implicit solvents (Fig. \ref{fig3}d) \cite{fattebert2002, arias2012, mathew2014,mathew2019}. These methods effortlessly describe the thermodynamic limit, i.e. do not suffer time-sampling limitations, but are currently unable to describe the local potential fluctuations. Furthermore, hybrid solvation schemes have been suggested (Fig. \ref{fig3}e), where a region of explicit electrolyte is complemented by a charged implicit solvent or continuum dielectric \cite{filhol2015,gross2015,goldsmith,Melander2024}.  These implicit solvents are placed at the edge of the explicit electrolyte region. More recently, QM/MM approaches such as DFT-CES (density functional theory with classical explicit solvents) \cite{lim2016, ringe2022} have extended traditional implicit solvents beyond (modified) Poisson-Boltzmann models. Hybrid approaches of this type require an implementation within the respective electronic structure package. 

Minimally invasive approaches working within the constraints of 3D PBC have been therefore exploited, as well. Multiple groups proposed \cite{gche1,gche2,agross2020,omranpoor2023} to vary the number of H$^+$ ions contained within the DFT supercell grand-canonically. For the same purpose, Cucinotta {\it et al.} \cite{cucinotta} introduced an intentionally unbalanced number of cations and anions in the electrolyte. These approaches give rise to an uncompensated net-charge. Since the total system is charge neutral, the non-compensated charge accumulates at the interface, cf. Fig. \ref{fig3}f. By construction, this charge can assume only integer values. Even when using rather large supercells, the number of possible charge states at the interface is limited. This limited number of charge states at the interface severely restricts the variability of the ensuing electric fields and, hence, the applicability of techniques to sample thermodynamically open boundary conditions with $\langle \Phi \rangle = \text{const}$.

For this reason, Surendralal {\it et al.} \cite{surendralal2018} proposed to replace the charged implicit or explicit region by an explicit atomistic counter electrode using pseudo-atoms, i.e. neutral objects with fractional proton numbers \cite{shiraishi}, to charge the working electrode. The employed setup introduces an explicit counter electrode consisting of Ne pseudo-atoms and a vacuum region into the supercell, and is shown in Fig.~\ref{fig3}g. Apart from enabling a continuous variation of the charge within the supercell, this approach has also the advantage, that: (i) it avoids the use of a compensating background, since the overall supercell remains neutral, (ii) all its design elements can be easily realized in a standard DFT code and do not require changes to the code and (iii) it works equally well with metallic and semiconducting electrodes. Moreover, Surendralal \emph{et al.}'s setup is straightforwardly extended to describe fully open boundary conditions for the electrode charge \cite{deissenbeck2023}. Dudzinski \emph{et al.} \cite{dudzinski2023} recently extended the idea of using fractionally charged pseudo-atoms to modify the total charge of the water molecules in an aqueous electrolyte, although at the price of relinquishing control over the distribution of charges within the electrolyte.

\begin{table}
\begin{centering}
\begin{minipage}{\linewidth}
\renewcommand{\thefootnote}{\themfootnote}
\newcommand\sbullet[1][.5]{\mathbin{\vcenter{\hbox{\scalebox{#1}{$\bullet$}}}}}
\begin{tabular}{l|c||c|c|c|c}
\hline
\multicolumn{2}{c}{Charging the supercell}  \&  &\multicolumn{4}{c}{Application to solid/liquid interfaces} \\
\hline
Method  & Charging & \multicolumn{2}{c|}{Geometry} & Boundary & Electrode\\
        &          & slab & solvent & conditions & potential \\
\hline \hline
MTP\footnote{Modern theory of polarization} \cite{pasquarello2002}, \cite{stengel2009}, \cite{anderssonlin} & continuous & sym & vacuum & const $E \oplus D$\footnote{$\oplus$: Exclusive or} &  no \\
\hline 

DC\footnote{Dipole correction} \cite{neugebauer1993} \& GDC\footnote{Generalized dipole correction}  \cite{freysoldt2020gdc} & continuous & sym & vacuum & const $E$ & yes \\
\hline 

Sheet charge/jellium \cite{lozovoi} & discrete & sym & vacuum & const $D$ & yes   \\
$\quad $
DRM\footnote{Double-reference method} \cite{filhol2006, taylor2006},& & & expl\,\footnote{explicit water} & & yes\\ 
\quad\quad\quad\quad\quad \cite{filhol2015}& & & impl\,\footnote{implicit solvent/water} & & yes\\ 
\hline

\multicolumn{6}{l}{Electrolyte with net charge}\\ 
$\quad $
VASPsol \cite{mathew2019, plaisance2023} & continuous & sym & impl\,\footnote{hybrid solvation, i.e. regions of explicit and implicit water simultaneously present, possible} & const $D$ & yes \\
$\quad $
Environ \cite{environ1,environ2,goldsmith} & continuous & asym & hybrid & const $D$ & yes\\
$\quad $
Explicit H$^+$ \cite{gche1,gche2,agross2020,omranpoor2023} & discrete & asym & expl & const $D$ & yes\\
$\quad $
Imbalanced ion conc. \cite{cucinotta} & discrete &  sym & expl & const $D$ & no\\
$\quad $
Fract. charged water \cite{dudzinski2023} & continuous & sym & expl & const $D$ &  no\\
$\quad $
ESM-RISM\footnote{Effective screening medium - Reference interaction site model} \cite{otani2006esm,otani2017,otani2022}& continuous &  asym & impl & const E & yes \\ 
\hline 

CIP-DFT\footnote{Constant inner potential DFT} \cite{Melander2024} & continuous & asym & hybrid & const $E$ & yes \\
\hline

Grand Canonical-DFT \cite{arias2017} & continuous & asym & impl & const $E$ & yes \\
\hline

Hairy Probes \cite{cucinotta2024} & continuous & asym & expl & const $E$ & yes \\
\hline

FCP\footnote{Fictitious charge particle method} \cite{otani2012PRL} \cite{bouzid2017} & continous &  asym & expl & const $\langle E \rangle$ & yes \\
\hline

CCE\footnote{Computational counter electrode} \cite{surendralal2018}, \cite{deissenbeck2021}, \cite{deissenbeck2023} & continuous & asym & expl & const $\langle E \rangle$ & yes \\
\hline 

FSE\footnote{Fully solvated electrode} \cite{deissenbeck2023} & continuous & asym & expl & const $\langle E \rangle$ & yes \\
\hline 
\end{tabular}
\end{minipage}
\caption{Methods to charge supercells and model electrochemical solid/liquid interfaces, and their key properties/approximations. We emphasize that the terms "constant electric field" (const E) and "constant displacement field" (const D) are common names for the underlying concepts \cite{resta, stengel2007, stengel2009, sprikPRB} and should not be taken literally, i.e., they do not imply a uniform local field or a homogeneous surface charge distribution at the electrode interface. Rather, they refer to the macroscopic laterally averaged boundary conditions imposed at the edges of the periodic simulation cell \cite{stengel2009}. The local fields and the surface charge distribution at the electrode interface can be highly inhomogeneous and are determined by the charge self-consistency in the DFT calculations when imposing the macroscopic laterally averaged boundary conditions.} Moreover, we note that methods enforcing a constant displacement field $D$ can, in principle, be adapted to impose a constant electric field $E$ (unless $D$ is constrained by system-specific factors, such as an integer number of compensating ions), but this generally requires an additional self-consistency loop.\label{table}
\end{centering}
\end{table}

We emphasize that modelling the electric double layer either as a charged continuum region outside the explicit electrolyte or as an object consisting of actual atoms creates a hydrophobic gap between the surrogate model and the electrolyte, leading to an undesired and non-negligible potential drop within this region \cite{deissenbeck2023}. Moreover, this interfacial potential loss at the surrogate model increases the likelihood of reaching a dielectric breakdown \cite{yoo2021} as the field strength increases (see also discussion in section \ref{sec:alignment}). 

To eliminate the hydrophobic gap at the computational counter electrode and any associated potential loss, in a newer development \cite{deissenbeck2023} the setup proposed by Surendralal {\it et al.} \cite{surendralal2018} has been extended to the use of a fully solvated electrode, cf. Fig.~\ref{fig3}h. In this approach a Gaussian counter charge, placed within the aqueous region, is used to charge the working electrode. To maintain the desired bulk water density, a confining force is applied to the water surface facing the periodic image of the working electrode. In the ensuing vacuum region, the dipole correction \cite{neugebauer1992adsorbate} is employed to remove any electric fields and eliminate spurious interactions between periodic images.  On the right side of the cell, the potential is given by the workfunction of the  uncharged (backside) surface of the slab facing the vacuum region (as in Fig.~\ref{fig3}e and g). This potential can be calculated for a free-standing uncharged slab in vacuum and its absolute value (workfunction) can be straightforwardly computed and used as an absolute potential reference. 

While this method requires slight modifications in the DFT code, it enables considerably larger potential drops and realistically strong electric fields at the working electrode/water interface. Moreover, this development allows to apply arbitrarily shaped countercharges. By determining the countercharge distribution self-consistently, such an approach could accurately describe the ionic charge distribution of the electrochemical double layer in that region, in order to realistically capture the diffuse region at the interface.

As a final point, we emphasize that in DFT simulations of electrochemical interfaces, only the near-surface region of the double layer is captured within the supercell, while the gradual potential decay toward the bulk electrolyte remains unresolved (cf. Fig. \ref{fig6}a). This gap can be bridged by applying \emph{a posteriori} corrections using physically motivated models of the double layer, such as Poisson–Boltzmann distributions or electrostatic models informed by the computed charge distribution near the interface. These approaches are conceptually similar to methods used to correct long-range electrostatic effects in charged supercells  \cite{defect_review}, where the external potential is approximated based on physical constraints. 

Table \ref{table} presents a summary of current methods used in DFT calculations to model electrochemical interfaces, highlighting their geometrical set-ups, charging capabilities, as well as their ability to enable electrode potential control. It also indicates the electrostatic boundary conditions commonly used within the respective schemes. These boundary conditions are discussed in detail in the following section.

\section{Electrostatic boundary conditions}
\label{sec:boundcon}

In the preceding section, we gave an overview about available surrogate models which replicate the impact of the electrochemical double layer and the real system's potential distribution inside the supercell. A critical aspect of such a construct is selecting appropriate electrostatic boundary conditions for the system, which have to include the working electrode, explicit electrolyte, and surrogate model. It is important to note that different regions of the electrochemical cell are governed by distinctly different time scales. The double layer relaxes slowly on nanosecond timescales due to the relatively slow ionic motion \cite{santos2024}, while the electrode surface equilibrates rapidly, on timescales governed by electron mobility \cite{melander_dyn}. The latter timescale is orders of magnitude faster. The chosen electrostatic boundary conditions must describe both the fast equilibration of the electrode charge and the slow relaxation of the double layer (via the surrogate model) equally well. This is particularly relevant under reactive conditions, where charge transfer processes must be captured, including the energetics and reaction rates of electrochemical processes.

\begin{figure}[t]%
\centering
\includegraphics[width=0.98\textwidth]{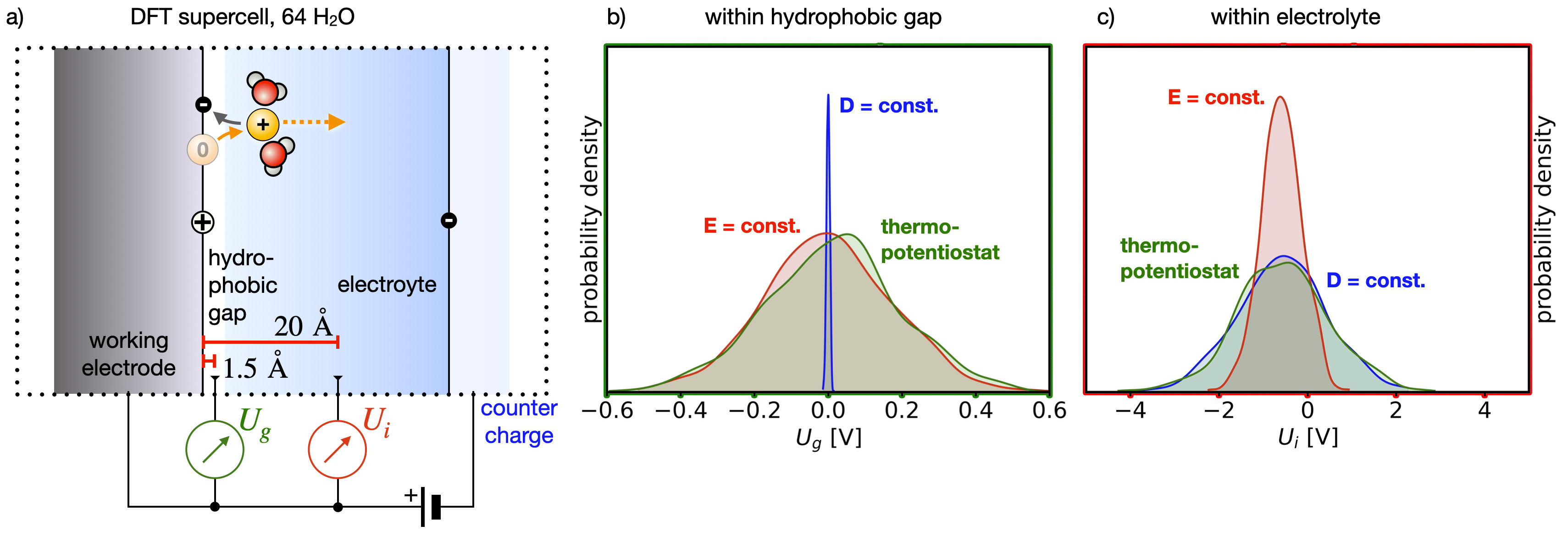}
\caption{\textbf{a)} Schematic representation of the supercell with an applied voltage between the working electrode and the surrogate double layer model. Voltages are measured between the working electrode and regions in close vicinity to the electrode surface ($U_g$) or deeper within the electrolyte region ($U_i$), respectively. \textbf{b/c)} Measured voltage distributions for different electrostatic boundary conditions \cite{flucts}, see text. The height of the delta peak in \textbf{b)} is rescaled for better visibility.}
\label{fig2}
\end{figure}

As a prototype example, we consider here an initially neutral atom or molecule at the interface (cf. Fig. \ref{fig2}a) that becomes positively charged after a reaction. Such processes are a crucial step in metal dissolution in contact with water (e.g., wet corrosion) \cite{marcus2009, Huang2020Mg, Yuwono2019} or in an electrocatalytic reaction \cite{rossmeisl, janik2009, laasonen, santos2022}. The driving forces for these reactions are the large solvation energy gains upon charging the solute and the strong electric fields at the interface that favor charge movement along the field. A product of the here considered prototype reaction, namely a single atom which is (singly) positively charged after the reaction, is an extra electron, which initially will stay on the surface. A key question when modeling such a reaction is what happens to this extra charge on the surface and are the boundary conditions adequate to reproduce the behaviour of the macroscopic system?

To explore the validity and performance of different boundary conditions, we performed MD simulations for a typical DFT sized supercell with lateral dimensions of 10 {\AA} by 10 \AA, containing 64 water molecules. We measured the voltage drop between the working electrode and regions close to the electrode surface ($U_g$), and deeper inside the electrolyte ($U_i$) \cite{flucts}, respectively, mimicking the placement of reference electrodes, cf. Fig. \ref{fig2}a. The resulting potential distributions provide a direct insight into the short-range electrostatics near the electrode and the long-range behavior in the electrolyte, and allow thereby testing the validity of the commonly discussed and applied boundary conditions under reactive conditions.

As a first scenario we consider constant charge conditions. In this case, the net charge on the electrode, where the reaction takes place, and the counter charge are kept fixed, and their sum is zero, i.e., the total setup of electrode and surrogate model is charge neutral. This scenario corresponds to constant electric displacement $D$ at the electrode-electrolyte interface. In the absence of charge transfer, the value of $D$ is directly related to the surface charge density $\sigma$ at the interface with $D = \sigma$.

Under constant charge conditions (constant displacement), the number of excess charges on the electrode surface (represented as a white circle with a plus) remains fixed. This constraint is reflected in the distribution of the voltage $U_g$ between the working electrode and the region directly above its surface. In the absence of charge transfer to or from the working electrode, the total surface charge is conserved. Consequently, the voltage drop across this region remains constant, leading to a delta distribution for $U_g$ (blue line, Fig. \ref{fig2}b).

Once the charge transfer reaction proceeds, the resulting extra electron remains localized on the electrode surface and cannot be compensated for under constant charge conditions. This unbalanced electron, in conjunction with the positively charged ion in the electrolyte, forms an electric dipole oriented along the surface normal. For a supercell geometry with periodic boundary conditions, this dipole is infinitely repeated laterally, giving rise to an effective capacitance and thus a large surface dipole. Consequently, the mean of the voltage distribution $U_g$ shifts substantially from its initial value. This shift disrupts the electrostatic conditions necessary for further reactions of the same type, effectively halting the process.

This behavior illustrates a limitation of constant charge boundary conditions when simulating reactive processes, where the interface dipole can change, e.g., due to electrochemical reactions, changes in the surface or adsorbate geometry, etc. In a real system, the electrode surface charge equilibrates rapidly on timescales governed by the electron mobility. Thus, the real system does not feature a delta distribution for $U_g$, but obeys a Gaussian (normal) distribution with a deviation of $k_B T / C$ \cite{deissenbeck2021, deissenbeck2023}, where $C$ is the effective (average) capacitance of the interface at the given mean potential. This capacitance characterizes the ability of the electrode to accommodate charge fluctuations and depends on both the interfacial geometry and the electronic structure. Therefore, approximating the surface charge as constant requires a careful analysis of the validity of this approximation for the specific study.

We note, however, that for computing equilibrium quantities of a surface in contact with an electrolyte, i.e. where changes in the interfacial dipole are small -- such as the potential of zero charge (PZC) or the interfacial capacitance -- the constant charge approach ($D =$ const) for the special condition of zero charge is well suited and has been successfully applied in several studies \cite{surendralal2021,pzc_jle, sundararaman2022}. The reason why these boundary conditions work well in this case is that in the absence of chemical reactions the surface structure remains unchanged and the PZC is characterized by the condition that no net charge is transferred to or away from the electrode, which is automatically satisfied by the condition \mbox{$q =$ const.}

Moreover, one should keep in mind that the double layer relaxes relatively slowly, on the order of up to nanoseconds, due to the sluggish motion of ions in the electrolyte \cite{santos2024}. While constant charge boundary conditions fail to accurately describe the dynamics of the voltage $U_g$ close to the electrode surface, they are well-suited for capturing the dynamics of the voltage $U_i$ within the double layer region of the cell through the surrogate model. As shown in Fig. \ref{fig2}c (blue curve), the thermal motion of the electrolyte induces polarization fluctuations, which are correctly reproduced under constant charge boundary conditions.

In order to lift the inability of constant displacement (charge) schemes to account for the charge dynamics of the working electrode under reaction conditions, a second approach suggests to keep the voltage between the working electrode and the surrogate model precisely constant, i.e., fluctuations in the applied voltage are not possible ($U \equiv \left<U\right>$). To keep the potential constant, charges can be exchanged (i.e. flow back and forward) with the potentiostat. Thus, for the model reaction considered here, once the extra electron is at the electrode, a charge $q$ will be transferred until the shift between initial and final voltage across the cell is eliminated. Therefore, these boundary conditions provide a realistic description of the electrochemical current that flows when keeping the voltage constant and charge transfer reactions take place. Experimentally, this would be measured as current-voltage (U-I) characteristic.

Constant electric field boundary conditions are particularly effective in capturing the fast charge dynamics of the working electrode surface. Unlike the delta distribution obtained for $U_g$ under constant charge conditions (Fig. \ref{fig2}b, blue curve), these boundary conditions yield the expected Gaussian distribution for $U_g$ (Fig. \ref{fig2}b, red curve), accurately reflecting thermal fluctuations near the electrode. However, constant electric field conditions enforce an instantaneous relaxation of the double layer, as the voltage between the working electrode and the surrogate model is held precisely constant at all times. This assumption deviates significantly from the behavior of the real system, where ionic relaxation occurs on much slower time scales. As a result, the voltage distribution $U_i$ within the electrolyte narrows considerably when approaching the double layer region (Fig. \ref{fig2}c, red curve). For DFT-sized supercells, this distribution is much narrower than the one obtained under constant charge conditions, which approximates the double layer as a static entity.

As expected, constant charge boundary conditions are better suited for describing the slow relaxation dynamics of the double layer, while constant field boundary conditions more effectively capture the rapid charge dynamics of the electrode surface. However, neither approach is capable of accurately representing both aspects simultaneously, failing to reconcile the short-range behavior near the electrode with the long-range behavior deeper inside the electrolyte.

We emphasize that the choice of boundary conditions can significantly influence the predicted reaction mechanisms and rates in electrochemical simulations. For typical supercell sizes used to model the electrochemical interface, the average voltage across the electrode electrolyte interface can fluctuate by approximately $\Delta \Phi \sim \pm 1$ V \cite{PhD_sudarsan,rossmeisl,spohr}, as illustrated in Figs. \ref{fig2}b/c. These fluctuations directly impact the reaction barrier and, consequently, the reaction rate.

The reaction rate depends on three distinct timescales: (i) the timescale of the reaction itself ($\tau_{\text{reaction}}$), (ii) the timescale of the electrostatic potential fluctuations ($\tau_\Phi$) due to the fast equilibration of the electrode charge, and (iii) the much slower timescale over which the double layer equilibrates ($\tau_{\text{eq}}$). The latter includes solvent and ionic reorganization, such as dielectric relaxation of water and ion migration, and sets the rate at which the electrostatic environment around the reacting species can adjust. If $\tau_\Phi \gg \tau_{\text{reaction}}$, the reaction occurs under a quasi-static potential, and the rate reflects the instantaneous barrier. Conversely, if $\tau_\Phi \ll \tau_{\text{reaction}}$, the system averages over the potential fluctuations during the reaction. When $\tau_\Phi \sim \tau_{\text{reaction}}$, the dynamics of the fluctuations must be explicitly considered, as the reaction rate becomes sensitive to the correlation between potential fluctuations and barrier crossing.

In general, the reaction rate $w_{\text{reaction}}$ can be expressed as
\begin{equation*}
w_{\text{reaction}} \propto \int_{-\infty}^\infty P\left( \Phi; \tau_{\Phi}, \tau_{\text{reaction}}, \tau_{\text{eq}} \right) \cdot e^{-\frac{\mathcal{E}_{\text{barrier}}\left[\Phi\right]}{k_B T}} d\Phi,
\end{equation*}
where $\mathcal{E}_{\text{barrier}}$ is the reaction barrier as a function of the instantaneous potential $\Phi$, and $P$ is the probability density of the system to adopt a specific electrostatic potential $\Phi$. This expression isolates the potential-dependent contribution to the reaction rate for simplicity. A full treatment would also require integration over molecular positions and momenta, which are omitted here. We emphasize that this timescale discussion is conceptual; accurate rate calculations require advanced sampling techniques such as umbrella sampling \cite{umbrella}, metadynamics \cite{metadyn1,metadyn2}, forward flux \cite{Allen2006,Allen2009}, or the blue moon ensemble \cite{bluemoon}. Importantly, the probability density $P$ -- and hence the predicted reaction kinetics -- depends on the chosen boundary conditions, as these influence both the magnitude of potential fluctuations and the timescales over which they evolve.

The most general boundary conditions for simulating electrochemical supercells are thermodynamically open, allowing for both charge transfer and dynamical fluctuations while maintaining voltage control. In this framework, the applied voltage is fixed only on average (e.g. $\left<\Phi\right> =$ const.), enabling fluctuations around the mean value. On short time scales, the electrode surface charge and the voltage drop across the interface can vary significantly, reflecting the interplay between the rapid dynamics of electrode charge equilibration and the slower relaxation processes of the electrochemical double layer. These boundary conditions aim to accurately capture the inherently dynamic nature of the electrochemical interface.

Fully open boundary conditions imply, however, that the system is able to perform external work $\Delta W = \Delta q \cdot \Delta \Phi$, where $\Delta q$ and $\Delta \Phi$ are the change in the charge and the potential, respectively. Under constant displacement or constant field boundary conditions, either $\Delta q$ or $\Delta \Phi$ is artificially enforced to be exactly zero, respectively, so that $\Delta W \equiv 0$. Under open boundary conditions, however, $\Delta W$ becomes finite. The finite $\Delta W$ permits the system  to ``borrow''  on short time scales energy from the ensuing charge and potential fluctuations. This mechanism is analogous to the energy exchange in thermostat theories, e.g. \emph{``[...] when the observables of interest are dependent on the fluctuations rather than on the averages [...].''} \cite{csvr}. The ability to perform work is crucial for restoring the applied potential to its target value after events such as charge transfer or changes in surface structure, thereby maintaining reactive conditions within the simulation. On the other hand, the finite $\Delta W$ implies that introducing potential control via relaxing $\Phi(t)$ towards its targeted mean $\langle \Phi \rangle$ dissipates energy.

In order to balance the energy dissipated due to potential control, one may be tempted to replenish the dissipated energy via an explicit thermostat. We emphasize, however, that a thermostat acts indiscriminately on all degrees of freedom, whereas a potentiostat is able to affect only those vibrational degrees of freedom that couple to a change in the ensemble's dipole moment parallel to the direction of the applied electric field. Draining energy from one set of degrees of freedom and subsequently returning it to another invariably leads to a spurious energy transfer between them. Such an approach cannot restore the system to equilibrium. On the contrary, it would constantly drive the system out of equilibrium \cite{deissenbeck2023}.

Thus, any open boundary potential control scheme must therefore simultaneously introduce a thermodynamically consistent distribution of charge and potential fluctuations. The fundamental relation between energy dissipation (to control, e. g., the temperature, pressure or potential) and fluctuations (to balance the dissipation in thermodynamic equilibrium) is provided by the fluctuation-dissipation theorem (FDT) \cite{callen}. Analogous to thermostats, where an equation of motion for the exchange of heat with an external bath is derived from the FDT and that are nowadays widely used in MD simulations, constructing a potentiostat involves deriving an equation of motion for the exchange of charge with an external voltage source \cite{otani2012PRL} via the FDT \cite{deissenbeck2021}.

These fully open boundary conditions \cite{otani2012PRL,deissenbeck2021} accurately reproduce in a supercell sized box the dynamic behavior of voltage and charge measured within the full macroscopic system. These conditions are thus able to describe all aspects of an electrochemical system including electrochemical reactions or dynamic changes in the surface or adsorbate structure. Computing the voltage distributions near the electrode surface ($U_g$) and deeper within the electrolyte ($U_i$) confirms the effectiveness of thermodynamically open boundary conditions (Fig. \ref{fig2}a/b, green curves). At short range, they replicate the behavior of constant field boundary conditions, accurately describing the fast dynamics of the electrode charge. At long range, they converge to the behavior of constant displacement boundary conditions, correctly modelling the slower relaxation of the double layer \cite{flucts}.

\section{Open thermodynamic boundary conditions}\label{sec:openBC}

The surrogate double layer models discussed in sections \ref{sec:Electrodes} and \ref{sec:boundcon} enable AIMD simulations at finite applied electric field for supercells that are charge-neutral in total, but with a varying number of electrons $n_{e^-}$ per supercell and thus, by extension, a varying charge $n$ on the working electrode. Based on these concepts, we now turn to discuss the realization of open thermodynamic boundary conditions with respect to the electrode charge $n$, 
as outlined in section \ref{sec:boundcon}.

Any such control scheme relies on a conjugate pair of extensive and intensive quantities - such as entropy and temperature (thermostat), or volume and pressure (barostat) - where the extensive quantity is used to drive the fluctuating intensive one to a desired target value\footnote{We note that in the case of thermostats, the entropy is difficult to control in simulations. There is, hence, a need to perform a Legendre transformation in order to control the temperature via the particle kinetic energies instead. For barostats, in contrast, the volume is straightforwardly adjusted. In that respect, the construction of a potentiostat is conceptually much closer to a barostat than to a thermostat.}. In the context of potential control, it is straightforwardly shown that 
the electrode charge $n$ and the electrode potential $\Phi$ -- as well as the displacement $D$ and the field $E$ \cite{landau} -- form such a pair of conjugate quantities \cite{deissenbeck2021}. Since the electrode is connected to an external reservoir of charge, constructing a potentiostat amounts to deriving an equation of motion for the extensive charge $n_{e^\text{-}}(t)$, i.e. the exchange of charge between the electrode at potential $\Phi(t)$ and the external reservoir held at $\Phi_0$.

This task is complicated by the fact that the potentiostat is intended to describe the exchange of charge with a reservoir not just in static calculations, but in the context of molecular dynamics at finite temperature. The thermal motion of the electrons and ions, as well as the exchange of charge between the electrode and the reservoir, create thermal fluctuations in the electric field and - by extension - in the potential of the local system $\Phi(t)$. In the physical system, the identical argument applies to the external environment surrounding the microscopic region targeted by our simulations. As a result, the DFT supercell effectively resides within a fluctuating electric field generated by the thermal motion of electrons and ions in its local external environment. This embedding reflects the realistic coupling between the simulated region and its surroundings, essential for accurately modeling electrochemical processes.

It is the role of the potentiostat to mimic the contact of the supercell with this external environment. If the reservoir of charge were to be held at an exactly constant potential without any thermal fluctuations (effectively corresponding to $T = 0$ K), the supercell would dissipate its thermal potential fluctuations into the external reservoir, thereby violating the 2$^{\rm nd}$ law of thermodynamics. Such an artificial net energy drain would constantly drive the simulation system out of equilibrium. In order to prevent such a spurious energy transfer, the electric field fluctuations within the supercell must be in thermodynamic equilibrium with the field fluctuations of the external environment at the boundary of the supercell. Therefore, the external reservoir of charge itself must have a finite temperature.

The first such approach was introduced in 2012 by Bonnet \emph{et al.} \cite{otani2012PRL}, who drove the exchange of charge between electrode and reservoir via a Nose-Hoover algorithm. The Nose-Hoover approach is known to perform well for reasonably ergodic degrees of freedom. The electrode charge, however, is only a single degree of freedom and thus requires Nose-Hoover chains at the cost of an increased number of parameters (Nose-masses) and associated extra tuning. Moreover, the absence of suitable surrogate models for the double layer in 2012 led Bonnet \emph{et al.} to construct their potentiostat for systems carrying a net excess charge. Hence, the Nose-Hoover potentiostat requires the derivative of the total energy with respect to the net excess charge as an input quantity. This quantity is notoriously difficult to obtain in DFT simulations. Those requirements proved to be serious obstacles in implementing the Nose-Hoover potentiostat into existing DFT codes.

A general approach toward control schemes is provided by the fluctuation-dissipation theorem \cite{callen}: Due to the exchange of particles or energy with the reservoir, the system is able to perform external work and dissipate energy. At the same time, the thermal fluctuations of the external environment return energy to the system. In thermodynamic equilibrium, the dissipation and the energy gain from the fluctuations are exactly equal, on average.

In the context of potential control, driving the system potential $\Phi(t)$ towards the external potential $\Phi_0$ is the dissipation. Formally, in an equivalent circuit approach, the dissipation can be represented as a resistance $R$ between the part of the working electrode located inside the supercell and its surrounding external environment. Physically, the resistance $R$ describes the effective surface resistance of the electrode to dampen the small charge fluctuations at the electrochemical interface that arise at the surface due to thermal atomic vibrations, (electro-) chemical reactions etc.  This damping is consistent with the fluctuation–dissipation theorem and the emergence of Johnson–Nyquist noise \cite{henkel16}. We emphasize that this is not a circuit-level resistance between electrodes. Much like the friction coefficient in a Langevin thermostat governs energy exchange between a particle and a thermal bath, the resistance $R$ controls the rate at which the electrode surface exchanges charge and potential fluctuations with its electrostatic environment.

This dissipation must then be balanced by the energy gained from the thermal potential fluctuations of the local external environment at the boundary of the supercell. In the equivalent circuit approach, the fluctuations are then given by the Johnson-Nyquist noise generated by the resistance $R$. It was shown in Ref. \cite{deissenbeck2021}, that the fluctuation-dissipation theorem in this case takes the form

\begin{equation*}
f dt = \underbrace{-\frac{1}{\tau_{\Phi}} (\Phi - \Phi_0)\ dt}_{dissipation} + \underbrace{\sqrt{\frac{2}{\tau_{\Phi}} \frac{k_B T}{C_0}}\ dW_t}_{fluctuation},
\end{equation*}
where $f$ is the rate of change of the electrode potential $\Phi$, $\tau_\Phi$ is a relaxation time constant (typically in the order of 100\,fs \cite{deissenbeck2023}), $dW_t$ is the time differential of the Wiener stochastic process and $C_0$ the capacitance of the bare electrodes in the absence of a dielectric, i.e. $C = \epsilon_r C_0$ \cite{deissenbeck2021, deissenbeck2023}. 

We emphasize that this energy balance amounts to zero, on average, only in thermodynamic equilibrium, i. e. if the temperatures of the external environment and the supercell are equal. If, e. g., the temperature of the supercell is above the one of the external environment, the dissipation dominates. On the other hand, if the temperature of the supercell is less than the one of the environment, the energy gain via the fluctuations exceeds the dissipation. Therefore, any viable approach to potential control invariably acts also as a thermostat.

Such a \emph{thermopotentiostat} was recently derived by Dei\ss enbeck \emph{et al.} \cite{deissenbeck2021,deissenbeck2023} via Ito integration \cite{gardiner} of the fluctuation-dissipation relation shown above. Thereby, the thermopotentiostat provides a direct expression for the electrode charge $n$ 
at each discrete MD time step, under open thermodynamic boundary conditions. In contrast to the $D = \text{const.}$ and $E = \text{const.}$ approaches, no artificial constraints on either the charge or potential dynamics are imposed \footnote{In the limit of setting the equivalent resistance $R$ (or by extension the relaxation time $\tau_\Phi$) to either zero or infinity, respectively, electrostatic boundary conditions of $\vec{E} = \text{const.}$ or $\vec{D} = \text{const.}$ are recovered.}. Instead, the thermopotentiostat is mathematically guaranteed to sample the charge and potential distributions depicted in Fig. \ref{fig2}, which maintain their averages even under charge transfer reactions. In addition, it no longer requires the derivative of the total energy with respect to the charge as an input quantity. Instead, $\Phi(t) - \Phi_0$ is the only mandatory input quantity. In computational setups such as the ones shown in Figs. \ref{fig3}e, g, h this potential difference can be conveniently measured via the total dipole moment - given, e.g., by the dipole correction \cite{neugebauer1992adsorbate} - along the normal direction of the electrode surface. Therefore, in conjunction with any of those computational setups the thermopotentiostat can be straightforwardly included in any electronic structure package.

From a practical standpoint, whether a potentiostat is necessary depends on both the system size and the nature of the process being studied. For typical DFT simulations of electrochemical interfaces -- often involving 100–1000 atoms and lateral cell dimensions of 1–2 nm -- finite-size effects remain significant. In such systems, local fluctuations in the electrostatic potential, which can influence activation barriers, reaction selectivity, or adsorption geometries, are suppressed because the simulation cell is too small to capture the full range of independent microstates present in a macroscopic environment. Just as thermostats and barostats correct for the finite size of a system in temperature- or pressure-controlled simulations, the potentiostat serves to remove finite-size artifacts in voltage-controlled conditions. For this reason, and because the computational overhead is negligible, we recommend potential control in any simulation involving charge transfer, applied bias, or voltage-sensitive processes. In contrast, for closed equilibrium systems where no charge exchange occurs and the goal is, e.g., structural or spectroscopic characterization, ensemble control may be safely omitted -- though the implications of the chosen boundary conditions should always be assessed. As the scale and realism of simulations continue to grow, careful consideration of ensemble control will remain essential for making physically meaningful predictions.

\section{Band structure perspective on electrochemical interfaces} \label{sec:alignment}

\begin{figure}[t]%
\centering
\includegraphics[width=0.98\textwidth]{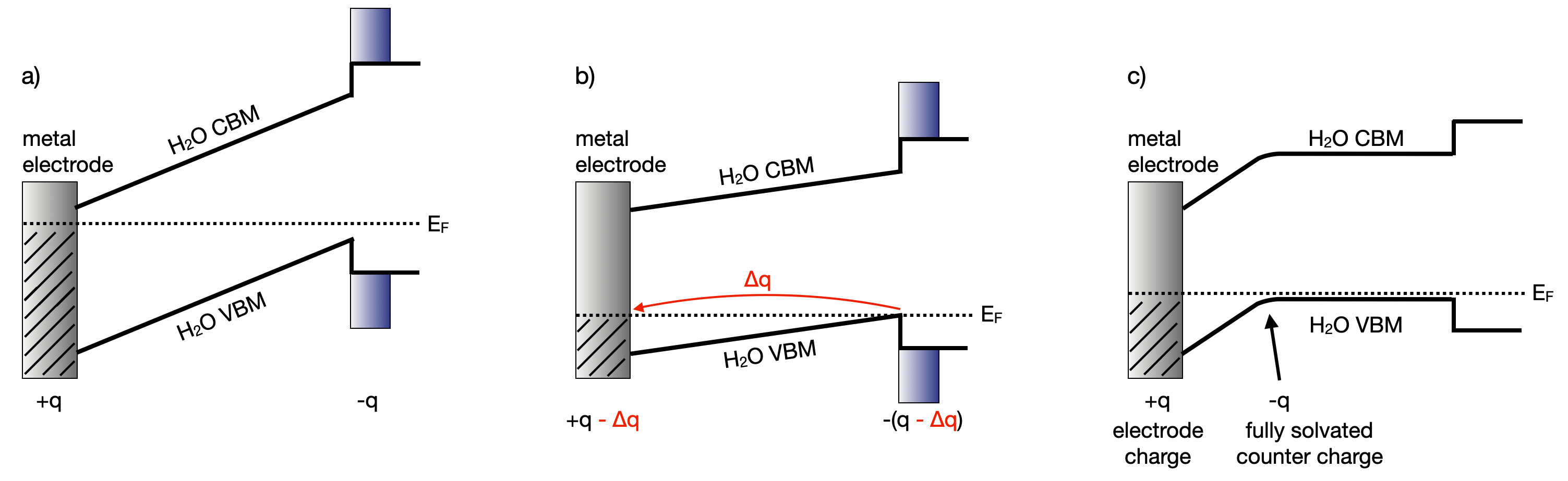}
\caption{\textbf{a)} Schematic representation of the band structure across an electrode-electrolyte interface. The field is applied via a semiconducting computational counter electrode (CCE) outside the electrolyte region. The Fermi level can be adjusted freely within the electronic gap of the electrolyte and the CCE. \textbf{b)} Sufficiently strong applied fields are able to tilt the bands of the electrolyte, so that the Fermi level straddles either the valence or conduction band edge. Dielectric breakdown will occur. A corresponding charge transfer between the working electrode and the CCE reduces the field so that the Fermi level is moved back to the band edge. \textbf{c)} Stronger fields can be applied at the same voltage without dielectric breakdown by moving the counter charge closer to the working electrode surface.}\label{fig4}
\end{figure}

While it is straightforward to apply electric fields in classical molecular dynamics \cite{deissenbeck2021}, special care must be taken in the context of electronic structure simulations. Fig. \ref{fig4}a shows a schematic representation of a prototypical interface under electric bias between a metal working electrode, explicit electrolyte (H$_2$O) and a surrogate model for the electric double layer. The surrogate model can be realized as discussed in the preceding section.

In any such setup, the electrolyte band gap defines the maximum possible potential drop, that can be applied across the electrolyte: if the applied voltage $U$ equals the band gap $E_g$, the energetic position of the conduction band minimum (CBM) at one side becomes equal to the energetic position of the valence band maximum (VBM) at the other side due to band bending. Fig. \ref{fig4}a depicts a situation where the applied voltage $U$ is only slightly less than $E_g$. For $U > E_g$, the VBM on the right hand side would be located energetically above the CBM on the left hand side, and, in turn, also above the Fermi energy. Hence, dielectric breakdown sets in, inducing a charge transfer $\delta q$, until the VBM shifts again below the Fermi energy. Therefore, the maximum electric field that can be achieved is $\|\vec{E}_{max}\| = E_g / d$, where $d$ is the thickness of the electrolyte.

We emphasize that in practical calculations, however, the maximum electric field that can be achieved before breakdown sets in is often much smaller than $E_g / d$: it also depends on the alignment of the Fermi level $E_F$ of the working electrode with respect to the band edges of the electrolyte at zero electrode charge (potential of zero charge: PZC). Although $E_F$ can be freely adjusted within $E_g$, dielectric breakdown equally occurs if the Fermi level coincides with either the VBM or CBM, respectively, at any location within the simulation cell due to the tilted bands. Such a situation is shown schematically in Fig. \ref{fig4}b: at the PZC, the Fermi level is located in close proximity to the VBM. Transferring a charge $q$ from the working electrode to the computational electrode - which is possible for the electronic structure shown in Fig. \ref{fig4}a - now lifts the VBM at the right hand side above $E_F$, causing dielectric breakdown. The charge $\Delta q$ contained in those valence states that are lifted above $E_F$ flows back into the working electrode, so that the VBM becomes equal to $E_F$: as a consequence, the Fermi level is pinned. The maximum critical field $\vec{E}_{max}$ that can be applied is, hence, the one just before Fermi level pinning occurs.

In any setup with explicit water
, the maximum electric field that can be applied depends on the thickness of the water slab contained within the simulation cell. At constant applied voltage, the total electric field decreases with increasing distance between the working and computational counter electrodes. For desirable cell sizes, these fields often fall short of realistic interfacial electric field strengths. The reason is that nature deposits the counter charges in the form of ions much closer to the electrode surface than possible via the computational counter electrodes depicted in Figs. \ref{fig4}a,b. Ref. \cite{deissenbeck2023} therefore suggested to decouple the position of the computational counter electrode from the edge of the explicit electrolyte region and place the counter charge directly within the electrolyte, cf. Fig. \ref{fig4}c. This allows achieving much stronger electric fields at the same applied voltage without causing dielectric breakdown. 

Here we emphasize that the situation is further complicated by the fact, that the applied electric field $\vec{E} = \varepsilon_0^{-1} [\vec{D} - \vec{P}]$ consists of two components: the equal and opposite charges $q$ on the working electrode and the computational counter electrode give rise to an electric displacement $\vec{D}$. In response to the electric displacement, the electrolyte becomes polarized with an ensuing polarization density $\vec{P}$. It is, however, their sum - the total field $\vec{E}$ - which is responsible for the band bending illustrated in Fig. \ref{fig4}. Accordingly, one may choose to work within different electrostatic boundary conditions of $\vec{D} = \text{const.}$ or $\vec{E} = \text{const.}$, as outlined in section \ref{sec:boundcon}.

As a consequence, it may be tempting to use $\vec{D} = \text{const.}$ and directly put the targeted charge on the working electrode. Unfortunately, the electrolyte is unable to reorganize instantaneously. For strongly polar solvents like water, in equilibrium $\vec{D}$ and $\vec{P}$ cancel out on average to a large degree, due to the large dielectric constant of water. The polarization density in the electrolyte is formed due to molecular reorientation and ion diffusion. These proceed on rather long time scales compared to those accessible in electronic structure simulations. Hence, polarizing a metal-electrolyte interface usually requires several tens of picoseconds \cite{deissenbeck2023}. Directly charging the working electrode under $\vec{D} = \text{const.}$ to its target value, without giving the electrolyte sufficient time to polarize, immediately causes dielectric breakdown, preventing in most cases even convergence of the DFT self-consistency loop.

One may argue that the $\vec{D} = \text{const.}$ approach may be applied safely by increasing the electrode charge $q$ only in sufficiently small increments, cf., e.g., Ref. \cite{surendralal2018}. This is true in principle. However, as discussed above, the electrons and ions undergo thermal fluctuations. By extension, their thermal movement creates fluctuations in the polarization density $\vec{P}(t)$, which lead to potential fluctuations on the order of $\sqrt{k_B T / C} \sim$ 1 V, where $C$ is the capacitance of the system. Even if $q$ is changed in only small increments, one has to check carefully and constantly that the polarization density fluctuations do not lead to electric fields $\vec{E}(t) = \varepsilon_0^{-1} [\vec{D} - \vec{P}(t)] > \vec{E}_{max}$.

Moreover, the system is driven out of thermodynamic equilibrium every time $q$ is changed incrementally. Changing $q$ too quickly may not only cause dielectric breakdown, but can also lead to simulation artefacts such as the flying-ice-cube-effect \cite{deissenbeck2023}. For all these reasons, boundary conditions of constant electric field $\vec{E} = \text{const.}$ are often preferred. As discussed in section \ref{sec:boundcon}, however, they are still unable to describe the correct thermodynamics under reactive conditions. Instead, it is preferable to go beyond $\vec{E} = \text{const.}$ and use fully open thermodynamic boundary conditions (Sec. \ref{sec:openBC}).

\section{Conclusion}

Electrochemical interfaces present a unique challenge for first-principles simulations, as they involve thermodynamically open systems that exchange energy, charge, and ions with their environment. These interfaces are governed by distinct timescales: the rapid charge equilibration of the electrode, driven by electron mobility; the slower ionic motion that defines the double-layer relaxation; and the intermediate timescales associated with electrochemical reactions. The interplay of these timescales leads to dynamic fluctuations in the local potential and electric field, which emerge from the complex, time-dependent interactions between electrode charge, solvent dynamics, and ionic distributions within the double layer.

These fluctuations are particularly relevant for electrochemical reactions, which are inherently local in time and space and therefore strongly influenced by transient potential fluctuations. Capturing these fluctuations is essential not only to replicate the correct mean potential but also to reproduce the thermal equilibrium behavior of the field. The challenge lies in ensuring that the finite DFT supercell reproduces the statistical and dynamic behavior of the effectively infinite electrochemical environment. Achieving this requires electrostatic boundary conditions that allow for the correct distribution of potential fluctuations, ensuring that reaction barriers, field distributions, and interfacial structures are realistically modeled—a task that remains challenging for conventional electronic structure methods.

In this review, we have analyzed the fundamental principles and available techniques for simulating electrochemical interfaces from first principles. We discussed the necessity of surrogate models for the electrochemical double layer and examined how different electrostatic boundary conditions influence potential distributions, reaction rates, and interfacial properties. We emphasized the importance of capturing not only the mean potential but also the fluctuations that characterize the system's thermal equilibrium, drawing analogies to the well-established role of thermostats in molecular dynamics.

A central finding is that boundary conditions enforcing either constant electric field or constant charge each capture only part of the interfacial physics. Constant field conditions reproduce fast electrode charge dynamics, while constant charge conditions better approximate the sluggish ionic relaxation of the double layer. Thermodynamically open boundary conditions provide a more complete framework by capturing both processes simultaneously. This approach enables simulations at constant potential on average - consistent with experimental conditions - while preserving the natural fluctuations and dynamics that govern reaction barriers and rates. Finally, we discussed issues related to band structure and band alignment, which need to be addressed to avoid unphysical simulation artifacts, e.g. by a dielectric breakdown or to understand limitations in the achievable maximum fields or applied voltages. 
 
While these methodological developments provide researchers with a new level of realism to study fundamental electrochemical mechanisms and reactions using \emph{ab initio} techniques, critical aspects remain open. An open conceptual challenge remains in the treatment of lateral charge dynamics. While the thermopotentiostat enables controlled charge exchange normal to the electrode surface, lateral directions typically assume instantaneous electron equilibration through periodic boundary conditions -- effectively enforcing constant electric field boundary conditions. This reflects the Born-Oppenheimer approximation, which neglects the finite lateral conductivity and surface resistance of real electrodes. In reality, lateral charge redistribution is governed by electron dynamics and finite surface mobility. Recent studies, for example in the context of surface contaminants in ion traps, have shown that finite surface resistance can give rise to lateral fluctuation-induced currents, which in turn affect the motion of nearby ions \cite{henkel16}. Such effects may become relevant in systems with spatially extended redox processes or fluctuating surface inhomogeneities. Incorporating explicit electron dynamics -- such as real-time TDDFT or Car-Parrinello-like approaches -- offers a promising route to go beyond the Born-Oppenheimer approximation and capture lateral equilibration processes, thereby enhancing the realism of potential-controlled simulations.

Moreover, these simulations are computationally extremely expensive, limiting fully \emph{ab initio} MD simulations with explicit water and electric fields to a few dozen picoseconds. This often limits our ability to get thermodynamically well-converged properties, explore rare events such as chemical reactions with high barriers, or to use improved DFT exchange-correlation functionals, e.g. hybrid functionals, which could help to overcome issues related to the severely underestimated bandgap of water when using semi-local functionals such as GGA. The recent success of machine learning potentials shows great potential to substantially speed up such calculations. However, in contrast to many of the existing approaches where short-range potentials with an interaction cutoff radius of a few Angstrom are sufficient, the inherent electrostatic nature of electrochemical interfaces makes the inclusion of the long-range Coulomb interactions mandatory \cite{dellago2023}. Alternatively, improved implicit solvation or microsolvation techniques \cite{plaisance2023} that include a limited amount of explicit solvent directly at the electrified interface are conceivable. Hybrid approaches, where machine learning or atomistic calculations \cite{ringe2022} are used to describe the dielectric response of the electrolyte to the electrified interface may also be promising to reach realistic time scales. Moreover, the computational demand can be reduced even further by advanced sampling techniques, such as, e.g., umbrella sampling or metadynamics \cite{metadyn1}, upsampling \cite{Jung2023}, as well as accelerated molecular dynamics approaches, such as forward flux sampling \cite{Allen2006,Allen2009}.
 
On another note, implementing grand-canonical boundary conditions not only with respect to the electrode charge, but also in chemical space has remained very challenging in the context of \emph{ab initio} molecular dynamics simulations. Achieving pH-control, in particular, will be needed. A possible route could be via hybrid QM/MM approaches, where hydronium ions are exchanged grand-canonically between the \emph{ab initio} simulation cell and the region modelled via, e.g., machine learning potentials or microsolvation.

Realizing these emerging techniques will provide access to more realistic time and length scales, as well as thermodynamically well-converged properties such as, e.g., reaction rates. Many electrochemical or -catalytical interfaces exhibit a multitude of possible reactions -- some desired, some detrimental -- with highly complex reaction networks. In turn, the developments outlined above will enable the exploration of these reaction networks and their critical mechanisms at first principles accuracy. The concepts and algorithms discussed in this review should be straightforwardly applicable to these extensions, opening a computationally efficient and physically consistent approach to address many of the still open issues related to electrochemical systems and reactions.

\backmatter

\bmhead{Acknowledgements}

The authors acknowledge the Deutsche Forschungsgemeinschaft (DFG, German Research Foundation) for funding through Projects No. 409476157 (SFB1394),  No. 506711657 (SFB1625), No. 223848855 (SFB1083) and support under Germany's Excellence Strategy - EXC 2033-390677874–RESOLV.  

\section*{Declarations}


\begin{itemize}
\item Funding: \\
DFG Project No. 409476157 (SFB1394), \\
DFG Project No. 506711657 (SFB1625), \\
DFG Project No. 223848855 (SFB1083).

\item Conflict of interest/Competing interests: 
There are no conflicts of interest or competing interests.

\item Ethics approval and consent to participate: 
Not applicable.

\item Consent for publication:
Not applicable.

\item Data availability: 
Not applicable.

\item Materials availability:
Not applicable.

\item Code availability: 
Not applicable.

\item Author contribution:
All authors contributed equally to this work. 
\end{itemize}

\noindent

%
%
%
%


\bibliography{sn-bibliography}

\begin{thebibliography}{100}
\expandafter\ifx\csname url\endcsname\relax
  \def\url#1{\burl{#1}}\fi
\expandafter\ifx\csname urlprefix\endcsname\relax\def\urlprefix{URL }\fi
\providecommand{\bibinfo}[2]{#2}
\providecommand{\eprint}[2][]{\url{#2}}
\providecommand{\doi}[1]{\url{https://doi.org/#1}}
\bibcommenthead

\bibitem{supercap1}
\bibinfo{author}{Wang, Y.}, \bibinfo{author}{Song, Y.} \& \bibinfo{author}{Xia,
  Y.}
\newblock \bibinfo{title}{Electrochemical capacitors: mechanism, materials,
  systems, characterization and applications}.
\newblock \emph{\bibinfo{journal}{Chem. Soc. Rev.}}
  \textbf{\bibinfo{volume}{45}}, \bibinfo{pages}{5925--5950}
  (\bibinfo{year}{2016}).

\bibitem{supercap2}
\bibinfo{author}{Wang, G.}, \bibinfo{author}{Zhang, L.} \&
  \bibinfo{author}{Zhang, J.}
\newblock \bibinfo{title}{A review of electrode materials for electrochemical
  supercapacitors}.
\newblock \emph{\bibinfo{journal}{Chem. Soc. Rev.}}
  \textbf{\bibinfo{volume}{41}}, \bibinfo{pages}{797--828}
  (\bibinfo{year}{2012}).

\bibitem{supercap3}
\bibinfo{author}{Zhu, Y.} \emph{et~al.}
\newblock \bibinfo{title}{Carbon-based supercapacitors produced by activation
  of graphene}.
\newblock \emph{\bibinfo{journal}{Science}} \textbf{\bibinfo{volume}{332}},
  \bibinfo{pages}{1537--1541} (\bibinfo{year}{2011}).

\bibitem{supercap4}
\bibinfo{author}{Simon, P.} \& \bibinfo{author}{Gogotsi, Y.}
\newblock \emph{\bibinfo{title}{Materials for electrochemical capacitors}},
  \bibinfo{pages}{320--329} (\bibinfo{publisher}{Co-Published with Macmillan
  Publishers Ltd, UK}, \bibinfo{year}{2009}).

\bibitem{supercap5}
\bibinfo{author}{Koetz, R.} \& \bibinfo{author}{Carlen, M.}
\newblock \bibinfo{title}{Principles and applications of electrochemical
  capacitors}.
\newblock \emph{\bibinfo{journal}{Electrochim. Acta}}
  \textbf{\bibinfo{volume}{45}}, \bibinfo{pages}{2483--2498}
  (\bibinfo{year}{2000}).

\bibitem{mab1}
\bibinfo{author}{Lee, J.-S.} \emph{et~al.}
\newblock \bibinfo{title}{Metal-air batteries with high energy density: Li-air
  versus {Z}n-air}.
\newblock \emph{\bibinfo{journal}{Adv. Energy Mater.}}
  \textbf{\bibinfo{volume}{1}}, \bibinfo{pages}{34--50} (\bibinfo{year}{2011}).

\bibitem{mab2}
\bibinfo{author}{Suntivich, J.} \emph{et~al.}
\newblock \bibinfo{title}{Design principles for oxygen-reduction activity on
  perovskite oxide catalysts for fuel cells and metal-air batteries}.
\newblock \emph{\bibinfo{journal}{Nat. Chem.}} \textbf{\bibinfo{volume}{3}},
  \bibinfo{pages}{546--550} (\bibinfo{year}{2011}).

\bibitem{mab3}
\bibinfo{author}{Cheng, F.} \& \bibinfo{author}{Chen, J.}
\newblock \bibinfo{title}{Metal-air batteries: from oxygen reduction
  electrochemistry to cathode catalysts}.
\newblock \emph{\bibinfo{journal}{Chem. Soc. Rev.}}
  \textbf{\bibinfo{volume}{41}}, \bibinfo{pages}{2172--2192}
  (\bibinfo{year}{2012}).

\bibitem{mab4}
\bibinfo{author}{Wang, Z.-L.}, \bibinfo{author}{Xu, D.}, \bibinfo{author}{Xu,
  J.-J.} \& \bibinfo{author}{Zhang, X.-B.}
\newblock \bibinfo{title}{Oxygen electrocatalysts in metal-air batteries: from
  aqueous to nonaqueous electrolytes}.
\newblock \emph{\bibinfo{journal}{Chem. Soc. Rev.}}
  \textbf{\bibinfo{volume}{43}}, \bibinfo{pages}{7746--7786}
  (\bibinfo{year}{2014}).

\bibitem{mab5}
\bibinfo{author}{Li, Y.} \& \bibinfo{author}{Lu, J.}
\newblock \bibinfo{title}{Metal air batteries: Will they be the future
  electrochemical energy storage device of choice?}
\newblock \emph{\bibinfo{journal}{ACS Energy Lett.}}
  \textbf{\bibinfo{volume}{2}}, \bibinfo{pages}{1370--1377}
  (\bibinfo{year}{2017}).

\bibitem{transient1}
\bibinfo{author}{Hwang, S.-W.} \emph{et~al.}
\newblock \bibinfo{title}{A physically transient form of silicon electronics}.
\newblock \emph{\bibinfo{journal}{Science}} \textbf{\bibinfo{volume}{337}},
  \bibinfo{pages}{1640--1644} (\bibinfo{year}{2012}).

\bibitem{transient2}
\bibinfo{author}{Yu, K.~J.} \emph{et~al.}
\newblock \bibinfo{title}{Bioresorbable silicon electronics for transient
  spatiotemporal mapping of electrical activity from the cerebral cortex}.
\newblock \emph{\bibinfo{journal}{Nat. Mater.}} \textbf{\bibinfo{volume}{15}},
  \bibinfo{pages}{782--791} (\bibinfo{year}{2016}).

\bibitem{transient3}
\bibinfo{author}{Ledesma, H.~A.} \& \bibinfo{author}{Tian, B.}
\newblock \bibinfo{title}{Nanoscale silicon for subcellular biointerfaces}.
\newblock \emph{\bibinfo{journal}{J. Mater. Chem. B}}
  \textbf{\bibinfo{volume}{5}}, \bibinfo{pages}{4276--4289}
  (\bibinfo{year}{2017}).

\bibitem{transient4}
\bibinfo{author}{Tian, B.} \emph{et~al.}
\newblock \bibinfo{title}{Roadmap on semiconductor-cell biointerfaces}.
\newblock \emph{\bibinfo{journal}{Phys. Bio.}} \textbf{\bibinfo{volume}{15}},
  \bibinfo{pages}{031002} (\bibinfo{year}{2018}).

\bibitem{plasma}
\bibinfo{author}{Sabat, K.}, \bibinfo{author}{Rajput, P.},
  \bibinfo{author}{Paramguru, R.}, \bibinfo{author}{Bhoi, B.} \&
  \bibinfo{author}{Mishra, B.}
\newblock \bibinfo{title}{Reduction of oxide minerals by hydrogen plasma: An
  overview}.
\newblock \emph{\bibinfo{journal}{Plasma Chem. Plasma Proc.}}
  \textbf{\bibinfo{volume}{34}}, \bibinfo{pages}{1--23} (\bibinfo{year}{2014}).

\bibitem{direct}
\bibinfo{author}{Cavaliere, P.}
\newblock \emph{\bibinfo{title}{Hydrogen Assisted Direct Reduction of Iron
  Oxides}}  (\bibinfo{publisher}{Springer Cham}, \bibinfo{year}{2022}).

\bibitem{reduction}
\bibinfo{author}{Lin, H.}, \bibinfo{author}{Chen, Y.} \& \bibinfo{author}{Li,
  C.}
\newblock \bibinfo{title}{The mechanism of reduction of iron oxide by
  hydrogen}.
\newblock \emph{\bibinfo{journal}{Thermochim. Acta}}
  \textbf{\bibinfo{volume}{400}}, \bibinfo{pages}{61--67}
  (\bibinfo{year}{2003}).

\bibitem{reduction2}
\bibinfo{author}{Spreitzer, D.} \& \bibinfo{author}{Schenk, J.}
\newblock \bibinfo{title}{Reduction of iron oxides with hydrogen - a review}.
\newblock \emph{\bibinfo{journal}{Steel Res. Int.}}
  \textbf{\bibinfo{volume}{90}}, \bibinfo{pages}{1900108}
  (\bibinfo{year}{2019}).

\bibitem{magnussen2019}
\bibinfo{author}{Magnussen, O.~M.} \& \bibinfo{author}{Groß, A.}
\newblock \bibinfo{title}{Toward an atomic-scale understanding of
  electrochemical interface structure and dynamics}.
\newblock \emph{\bibinfo{journal}{J. Am. Chem. Soc.}}
  \textbf{\bibinfo{volume}{141}}, \bibinfo{pages}{4777--4790}
  (\bibinfo{year}{2019}).

\bibitem{nose1}
\bibinfo{author}{Nose, S.}
\newblock \bibinfo{title}{A molecular dynamics method for simulations in the
  canonical ensemble}.
\newblock \emph{\bibinfo{journal}{Molec. Phys.}} \textbf{\bibinfo{volume}{52}},
  \bibinfo{pages}{255} (\bibinfo{year}{1984}).

\bibitem{nose2}
\bibinfo{author}{Nose, S.}
\newblock \bibinfo{title}{A unified formulation of the constant temperature
  molecular dynamics methods}.
\newblock \emph{\bibinfo{journal}{J. Chem. Phys.}}
  \textbf{\bibinfo{volume}{81}}, \bibinfo{pages}{511} (\bibinfo{year}{1984}).

\bibitem{berendsen}
\bibinfo{author}{Berendsen, H.}, \bibinfo{author}{Postma, J.},
  \bibinfo{author}{van Gunsteren, W.}, \bibinfo{author}{DiNola, A.} \&
  \bibinfo{author}{Haak, J.}
\newblock \bibinfo{title}{Molecular dynamics with coupling to an external
  bath}.
\newblock \emph{\bibinfo{journal}{J. Chem. Phys.}}
  \textbf{\bibinfo{volume}{81}}, \bibinfo{pages}{3684} (\bibinfo{year}{1984}).

\bibitem{nhchain}
\bibinfo{author}{Martyna, G.~J.}, \bibinfo{author}{Klein, M.} \&
  \bibinfo{author}{Tuckerman, M.}
\newblock \bibinfo{title}{Nosé–hoover chains: The canonical ensemble via
  continuous dynamics}.
\newblock \emph{\bibinfo{journal}{J. Chem. Phys.}}
  \textbf{\bibinfo{volume}{97}}, \bibinfo{pages}{2635} (\bibinfo{year}{1992}).

\bibitem{bussi1}
\bibinfo{author}{Bussi, G.}, \bibinfo{author}{Donadio, D.} \&
  \bibinfo{author}{Parinello, M.}
\newblock \bibinfo{title}{Canonical sampling through velocity rescaling}.
\newblock \emph{\bibinfo{journal}{J. Chem. Phys.}}
  \textbf{\bibinfo{volume}{126}}, \bibinfo{pages}{014101}
  (\bibinfo{year}{2007}).

\bibitem{bussi2}
\bibinfo{author}{Bussi, G.} \& \bibinfo{author}{Parinello, M.}
\newblock \bibinfo{title}{Stochastic thermostats: comparison of local and
  global schemes}.
\newblock \emph{\bibinfo{journal}{Comput. Phys. Comm.}}
  \textbf{\bibinfo{volume}{179}}, \bibinfo{pages}{26--29}
  (\bibinfo{year}{2008}).

\bibitem{callen}
\bibinfo{author}{Callen, H.~B.} \& \bibinfo{author}{Welton, T.~A.}
\newblock \bibinfo{title}{Irreversibility and generalized noise}.
\newblock \emph{\bibinfo{journal}{Phys. Rev.}} \textbf{\bibinfo{volume}{83}},
  \bibinfo{pages}{34} (\bibinfo{year}{1951}).

\bibitem{amatore2004}
\bibinfo{author}{Grün, F.}, \bibinfo{author}{Jardat, M.},
  \bibinfo{author}{Turq, P.} \& \bibinfo{author}{Amatore, C.}
\newblock \bibinfo{title}{Relaxation of the electrical double layer after an
  electron transfer approached by brownian dynamics simulation}.
\newblock \emph{\bibinfo{journal}{J. Chem. Phys.}}
  \textbf{\bibinfo{volume}{120}}, \bibinfo{pages}{9648–9655}
  (\bibinfo{year}{2004}).

\bibitem{santos2024}
\bibinfo{author}{Santos, E.}, \bibinfo{author}{Aradi, B.},
  \bibinfo{author}{van~der Heide, T.} \& \bibinfo{author}{Schmickler, W.}
\newblock \bibinfo{title}{Free energy curves for the volmer reaction obtained
  from molecular dynamics simulation based on quantum chemistry}.
\newblock \emph{\bibinfo{journal}{J. Electroanal. Chem.}}
  \textbf{\bibinfo{volume}{954}}, \bibinfo{pages}{118044}
  (\bibinfo{year}{2024}).

\bibitem{rev_reuter}
\bibinfo{author}{Ringe, S.}, \bibinfo{author}{Hörmann, N.},
  \bibinfo{author}{Oberhofer, H.} \& \bibinfo{author}{Reuter, K.}
\newblock \bibinfo{title}{Implicit solvation methods for catalysis at
  electrified interfaces}.
\newblock \emph{\bibinfo{journal}{Chem. Rev.}} \textbf{\bibinfo{volume}{122}},
  \bibinfo{pages}{10777–10820} (\bibinfo{year}{2021}).

\bibitem{schwarz2020}
\bibinfo{author}{Schwarz, K.} \& \bibinfo{author}{Sundararaman, R.}
\newblock \bibinfo{title}{The electrochemical interface in first-principles
  calculation}.
\newblock \emph{\bibinfo{journal}{Surf. Sci. Rep.}}
  \textbf{\bibinfo{volume}{75}}, \bibinfo{pages}{100492}
  (\bibinfo{year}{2020}).

\bibitem{rev_gross}
\bibinfo{author}{Gro\ss, A.} \& \bibinfo{author}{Sakong, S.}
\newblock \bibinfo{title}{Modelling the electric double layer at
  electrode/electrolyte interfaces}.
\newblock \emph{\bibinfo{journal}{Curr. Op. Electrochem.}}
  \textbf{\bibinfo{volume}{14}}, \bibinfo{pages}{1--6} (\bibinfo{year}{2019}).

\bibitem{rondinelly2019}
\bibinfo{author}{Huang, L.-F.}, \bibinfo{author}{Scully, J.~R.} \&
  \bibinfo{author}{Rondinelli, J.~M.}
\newblock \bibinfo{title}{Modeling corrosion with first-principles
  electrochemical phase diagrams}.
\newblock \emph{\bibinfo{journal}{Ann. Rev. Mater. Res.}}
  \textbf{\bibinfo{volume}{49}}, \bibinfo{pages}{53--77}
  (\bibinfo{year}{2019}).

\bibitem{koper2012}
\bibinfo{author}{Calle-Vallejo, F.} \& \bibinfo{author}{Koper, M.~T.}
\newblock \bibinfo{title}{First-principles computational electrochemistry:
  Achievements and challenges}.
\newblock \emph{\bibinfo{journal}{Electrochim. Acta}}
  \textbf{\bibinfo{volume}{84}}, \bibinfo{pages}{3--11} (\bibinfo{year}{2012}).

\bibitem{alfonso2018}
\bibinfo{author}{Alfonso, D.~R.}, \bibinfo{author}{Tafen, D.~N.} \&
  \bibinfo{author}{Kauffmann, D.~R.}
\newblock \bibinfo{title}{First-principles modeling in heterogeneous
  electrocatalysis}.
\newblock \emph{\bibinfo{journal}{Catal.}} \textbf{\bibinfo{volume}{8}},
  \bibinfo{pages}{424} (\bibinfo{year}{2018}).

\bibitem{Pastor2024}
\bibinfo{author}{Pastor, E.} \emph{et~al.}
\newblock \bibinfo{title}{Complementary probes for the electrochemical
  interface}.
\newblock \emph{\bibinfo{journal}{Nat. Rev. Chem.}}
  \textbf{\bibinfo{volume}{8}}, \bibinfo{pages}{159} (\bibinfo{year}{2024}).

\bibitem{butler2019}
\bibinfo{author}{Butler, K.~T.}, \bibinfo{author}{Gautam, G.~S.} \&
  \bibinfo{author}{Canepa, P.}
\newblock \bibinfo{title}{Designing interfaces in energy materials applications
  with first-principles calculations}.
\newblock \emph{\bibinfo{journal}{npj Comput. Mater.}}
  \textbf{\bibinfo{volume}{5}}, \bibinfo{pages}{19} (\bibinfo{year}{2019}).

\bibitem{rev_gonella}
\bibinfo{author}{Gonella, G.} \emph{et~al.}
\newblock \bibinfo{title}{Water at charged interfaces}.
\newblock \emph{\bibinfo{journal}{Nat. Rev. Chem.}}
  \textbf{\bibinfo{volume}{5}}, \bibinfo{pages}{466} (\bibinfo{year}{2021}).

\bibitem{che1}
\bibinfo{author}{Norskov, J.~K.} \emph{et~al.}
\newblock \bibinfo{title}{Origin of the overpotential for oxygen reduction at a
  fuel-cell cathode}.
\newblock \emph{\bibinfo{journal}{J. Phys. Chem. B}}
  \textbf{\bibinfo{volume}{108}}, \bibinfo{pages}{17886}
  (\bibinfo{year}{2004}).

\bibitem{filhol2006}
\bibinfo{author}{Filhol, J.-S.} \& \bibinfo{author}{Neurock, M.}
\newblock \bibinfo{title}{Elucidation of the electrochemical activation of
  water over {P}d by first principles}.
\newblock \emph{\bibinfo{journal}{Angew. Chem. Int. Ed.}}
  \textbf{\bibinfo{volume}{45}}, \bibinfo{pages}{402--406}
  (\bibinfo{year}{2006}).

\bibitem{taylor2006}
\bibinfo{author}{Taylor, C.~D.}, \bibinfo{author}{Wasileski, S.~A.},
  \bibinfo{author}{Filhol, J.-S.} \& \bibinfo{author}{Neurock, M.}
\newblock \bibinfo{title}{First principles reaction modeling of the
  electrochemical interface: Consideration and calculation of a tunable surface
  potential from atomic and electronic structure}.
\newblock \emph{\bibinfo{journal}{Phys. Rev. B}} \textbf{\bibinfo{volume}{73}},
  \bibinfo{pages}{165402} (\bibinfo{year}{2006}).

\bibitem{schnur}
\bibinfo{author}{Schnur, S.} \& \bibinfo{author}{Gross, A.}
\newblock \bibinfo{title}{Challenges in the first-principles description of
  reactions in electrocatalysis}.
\newblock \emph{\bibinfo{journal}{Catal. Today}}
  \textbf{\bibinfo{volume}{165}}, \bibinfo{pages}{129} (\bibinfo{year}{2011}).

\bibitem{gaigeot}
\bibinfo{author}{Sulpizi, M.}, \bibinfo{author}{Gaigeot, M.-P.} \&
  \bibinfo{author}{Sprik, M.}
\newblock \bibinfo{title}{The silica-water interface: How the silanols
  determine the surface acidity and modulate the water properties}.
\newblock \emph{\bibinfo{journal}{J. Chem. Theo. Comput.}}
  \textbf{\bibinfo{volume}{8}}, \bibinfo{pages}{1037--1047}
  (\bibinfo{year}{2012}).

\bibitem{Galli2021}
\bibinfo{author}{Ye, Z.}, \bibinfo{author}{Prominski, A.},
  \bibinfo{author}{Tian, B.} \& \bibinfo{author}{Galli, G.}
\newblock \bibinfo{title}{Probing the electronic properties of the electrified
  silicon/water interface by combining simulations and experiments}.
\newblock \emph{\bibinfo{journal}{Proc. Nat. Acad. Sci.}}
  \textbf{\bibinfo{volume}{118}}, \bibinfo{pages}{e2114929118}
  (\bibinfo{year}{2021}).

\bibitem{cyli}
\bibinfo{author}{Li, C.-Y.} \emph{et~al.}
\newblock \bibinfo{title}{In situ probing electrified interfacial water
  structures at atomically flat surfaces}.
\newblock \emph{\bibinfo{journal}{Nat. Mater.}} \textbf{\bibinfo{volume}{18}},
  \bibinfo{pages}{697} (\bibinfo{year}{2019}).

\bibitem{neugebauer1993}
\bibinfo{author}{Neugebauer, J.} \& \bibinfo{author}{Scheffler, M.}
\newblock \bibinfo{title}{Theory of adsorption and desorption in high electric
  fields}.
\newblock \emph{\bibinfo{journal}{Surf. Sci.}} \textbf{\bibinfo{volume}{287}},
  \bibinfo{pages}{572--576} (\bibinfo{year}{1993}).

\bibitem{neugebauer1992adsorbate}
\bibinfo{author}{Neugebauer, J.} \& \bibinfo{author}{Scheffler, M.}
\newblock \bibinfo{title}{Adsorbate-substrate and adsorbate-adsorbate
  interactions of {N}a and {K} adlayers on {A}l(111)}.
\newblock \emph{\bibinfo{journal}{Phys. Rev. B}} \textbf{\bibinfo{volume}{46}},
  \bibinfo{pages}{16067} (\bibinfo{year}{1992}).

\bibitem{lozovoi}
\bibinfo{author}{Lozovoi, A.}, \bibinfo{author}{Alavi, A.},
  \bibinfo{author}{Kohanoff, J.} \& \bibinfo{author}{Lynden-Bell, R.}
\newblock \bibinfo{title}{Ab initio simulation of charged slabs at constant
  chemical potential}.
\newblock \emph{\bibinfo{journal}{J. Chem. Phys.}}
  \textbf{\bibinfo{volume}{115}}, \bibinfo{pages}{1661} (\bibinfo{year}{2001}).

\bibitem{defect_review}
\bibinfo{author}{Freysoldt, C.} \emph{et~al.}
\newblock \bibinfo{title}{First-principles calculations for point defects in
  solids}.
\newblock \emph{\bibinfo{journal}{Rev. Mod. Phys.}}
  \textbf{\bibinfo{volume}{86}}, \bibinfo{pages}{253--305}
  (\bibinfo{year}{2014}).

\bibitem{FNV_PRL}
\bibinfo{author}{Freysoldt, C.}, \bibinfo{author}{Neugebauer, J.} \&
  \bibinfo{author}{Van~de Walle, C.~G.}
\newblock \bibinfo{title}{Fully ab initio finite-size corrections for
  charged-defect supercell calculations}.
\newblock \emph{\bibinfo{journal}{Phys. Rev. Lett.}}
  \textbf{\bibinfo{volume}{102}}, \bibinfo{pages}{016402}
  (\bibinfo{year}{2009}).

\bibitem{otani2006esm}
\bibinfo{author}{Otani, M.} \& \bibinfo{author}{Sugino, O.}
\newblock \bibinfo{title}{First-principles calculations of charged surfaces and
  interfaces: A plane-wave nonrepeated slab approach}.
\newblock \emph{\bibinfo{journal}{Phys. Rev. B}} \textbf{\bibinfo{volume}{73}},
  \bibinfo{pages}{115407} (\bibinfo{year}{2006}).

\bibitem{greens}
\bibinfo{author}{Hamada, I.}, \bibinfo{author}{Otani, M.},
  \bibinfo{author}{Sugino, O.} \& \bibinfo{author}{Morikawa, Y.}
\newblock \bibinfo{title}{Green’s function method for elimination of the
  spurious multipole interaction in the surface/interface slab model}.
\newblock \emph{\bibinfo{journal}{Phys. Rev. B}} \textbf{\bibinfo{volume}{80}},
  \bibinfo{pages}{165411} (\bibinfo{year}{2009}).

\bibitem{otani2013esm2}
\bibinfo{author}{Hamada, I.}, \bibinfo{author}{Sugino, O.},
  \bibinfo{author}{Bonnet, N.} \& \bibinfo{author}{Otani, M.}
\newblock \bibinfo{title}{Improved modeling of electrified interfaces using the
  effective screening medium method}.
\newblock \emph{\bibinfo{journal}{Phys. Rev. B}} \textbf{\bibinfo{volume}{88}},
  \bibinfo{pages}{155427} (\bibinfo{year}{2013}).

\bibitem{pasquarello2002}
\bibinfo{author}{Umari, P.} \& \bibinfo{author}{Pasquarello, A.}
\newblock \bibinfo{title}{Ab initio molecular dynamics in a finite homogeneous
  electric field}.
\newblock \emph{\bibinfo{journal}{Phys. Rev. Lett.}}
  \textbf{\bibinfo{volume}{89}}, \bibinfo{pages}{157602}
  (\bibinfo{year}{2002}).

\bibitem{resta}
\bibinfo{author}{Resta, R.}
\newblock \bibinfo{title}{Macroscopic polarization in crystalline dielectrics:
  the geometric phase approach}.
\newblock \emph{\bibinfo{journal}{Rev. Mod. Phys.}}
  \textbf{\bibinfo{volume}{66}}, \bibinfo{pages}{899} (\bibinfo{year}{1994}).

\bibitem{stengel2007}
\bibinfo{author}{Stengel, M.} \& \bibinfo{author}{Spaldin, N.~A.}
\newblock \bibinfo{title}{Ab initio theory of metal-insulator interfaces in a
  finite electric field}.
\newblock \emph{\bibinfo{journal}{Phys. Rev. B}} \textbf{\bibinfo{volume}{75}},
  \bibinfo{pages}{205121} (\bibinfo{year}{2007}).

\bibitem{stengel2009}
\bibinfo{author}{Stengel, M.}, \bibinfo{author}{Spaldin, N.~A.} \&
  \bibinfo{author}{Vanderbilt, D.}
\newblock \bibinfo{title}{Electric displacement as the fundamental variable in
  electronic-structure calculations}.
\newblock \emph{\bibinfo{journal}{Nat. Phys.}} \textbf{\bibinfo{volume}{5}},
  \bibinfo{pages}{304--308} (\bibinfo{year}{2009}).

\bibitem{sandia}
\bibinfo{title}{Publicly available {DFT} codes:
  https://dft.sandia.gov/quest/dft\_codes.html}
  \urlprefix\url{https://dft.sandia.gov/Quest/DFT_codes.html}.

\bibitem{marzari1997}
\bibinfo{author}{Marzari, N.} \& \bibinfo{author}{Vanderbilt, D.}
\newblock \bibinfo{title}{Maximally localized generalized wannier functions for
  composite energy bands}.
\newblock \emph{\bibinfo{journal}{Phys. Rev. B}} \textbf{\bibinfo{volume}{56}},
  \bibinfo{pages}{12847} (\bibinfo{year}{1997}).

\bibitem{fattebert2002}
\bibinfo{author}{Fattebert, J.-L.} \& \bibinfo{author}{Gygi, F.}
\newblock \bibinfo{title}{First‐principles molecular dynamics simulations in
  a continuum solvent}.
\newblock \emph{\bibinfo{journal}{Int. J. Quantum Chem.}}
  \textbf{\bibinfo{volume}{93}}, \bibinfo{pages}{139--147}
  (\bibinfo{year}{2003}).

\bibitem{arias2012}
\bibinfo{author}{Letchworth-Weaver, K.} \& \bibinfo{author}{Arias, T.~A.}
\newblock \bibinfo{title}{Joint density functional theory of the
  electrode-electrolyte interface: Application to fixed electrode potentials,
  interfacial capacitances, and potentials of zero charge}.
\newblock \emph{\bibinfo{journal}{Phys. Rev. B}} \textbf{\bibinfo{volume}{86}},
  \bibinfo{pages}{075140} (\bibinfo{year}{2012}).

\bibitem{mathew2014}
\bibinfo{author}{Mathew, K.}, \bibinfo{author}{Sundararaman, R.},
  \bibinfo{author}{Letchworth-Weaver, K.}, \bibinfo{author}{Arias, T.} \&
  \bibinfo{author}{Hennig, R.~G.}
\newblock \bibinfo{title}{Implicit solvation model for density-functional study
  of nanocrystal surfaces and reaction pathways}.
\newblock \emph{\bibinfo{journal}{J. Chem. Phys.}}
  \textbf{\bibinfo{volume}{140}}, \bibinfo{pages}{084106}
  (\bibinfo{year}{2014}).

\bibitem{mathew2019}
\bibinfo{author}{Mathew, K.}, \bibinfo{author}{Kolluru, V. S.~C.},
  \bibinfo{author}{Mula, S.}, \bibinfo{author}{Steinmann, S.~N.} \&
  \bibinfo{author}{Hennig, R.~G.}
\newblock \bibinfo{title}{Implicit self-consistent electrolyte model in
  plane-wave density-functional theory}.
\newblock \emph{\bibinfo{journal}{J. Chem. Phys.}}
  \textbf{\bibinfo{volume}{151}}, \bibinfo{pages}{234101}
  (\bibinfo{year}{2019}).

\bibitem{filhol2015}
\bibinfo{author}{Lespes, N.} \& \bibinfo{author}{Filho, J.-S.}
\newblock \bibinfo{title}{Using implicit solvent in ab initio electrochemical
  modeling: Investigating {L}i$^+$/{L}i electrochemistry at a {L}i/solvent
  interface}.
\newblock \emph{\bibinfo{journal}{J. Chem.Theory Comput.}}
  \textbf{\bibinfo{volume}{11}}, \bibinfo{pages}{3375 -- 3382}
  (\bibinfo{year}{2015}).

\bibitem{gross2015}
\bibinfo{author}{Sakong, S.}, \bibinfo{author}{Naderian, M.},
  \bibinfo{author}{Mathew, K.}, \bibinfo{author}{Hennig, R.~G.} \&
  \bibinfo{author}{Groß, A.}
\newblock \bibinfo{title}{Density functional theory study of the
  electrochemical interface between a {P}t electrode and an aqueous electrolyte
  using an implicit solvent method}.
\newblock \emph{\bibinfo{journal}{J. Chem. Phys.}}
  \textbf{\bibinfo{volume}{142}}, \bibinfo{pages}{234107}
  (\bibinfo{year}{2015}).

\bibitem{goldsmith}
\bibinfo{author}{Goldsmith, Z.~K.}, \bibinfo{author}{Calegari~Andrade, M.~F.}
  \& \bibinfo{author}{Selloni, A.}
\newblock \bibinfo{title}{Effects of applied voltage on water at a gold
  electrode interface from ab initio molecular dynamics}.
\newblock \emph{\bibinfo{journal}{Chem. Sci.}} \textbf{\bibinfo{volume}{12}},
  \bibinfo{pages}{5865} (\bibinfo{year}{2021}).

\bibitem{Melander2024}
\bibinfo{author}{Melander, M.~M.}, \bibinfo{author}{Wu, T.},
  \bibinfo{author}{Weckman, T.} \& \bibinfo{author}{Honkala, K.}
\newblock \bibinfo{title}{Constant inner potential {DFT} for modelling
  electrochemical systems under constant potential and bias}.
\newblock \emph{\bibinfo{journal}{npj Comput. Mater.}}
  \textbf{\bibinfo{volume}{10}}, \bibinfo{pages}{5} (\bibinfo{year}{2024}).

\bibitem{lim2016}
\bibinfo{author}{Lim, H.-K.}, \bibinfo{author}{Lee, H.} \&
  \bibinfo{author}{Kim, H.}
\newblock \bibinfo{title}{A seamless grid-based interface for mean-field
  {QM}/{MM} coupled with efficient solvation free energy calculations}.
\newblock \emph{\bibinfo{journal}{J. Chem. Theory Comput.}}
  \textbf{\bibinfo{volume}{12}}, \bibinfo{pages}{5088–5099}
  (\bibinfo{year}{2016}).

\bibitem{ringe2022}
\bibinfo{author}{Shin, S.-J.} \emph{et~al.}
\newblock \bibinfo{title}{On the importance of the electric double layer
  structure in aqueous electrocatalysis}.
\newblock \emph{\bibinfo{journal}{Nat. Commun.}} \textbf{\bibinfo{volume}{13}},
  \bibinfo{pages}{174} (\bibinfo{year}{2022}).

\bibitem{gche1}
\bibinfo{author}{Hansen, M.} \& \bibinfo{author}{Rossmeisl, J.}
\newblock \bibinfo{title}{Finite bias calculations to model interface dipoles
  in electrochemical cells at the atomic scale}.
\newblock \emph{\bibinfo{journal}{J. Phys. Chem. C}}
  \textbf{\bibinfo{volume}{120}}, \bibinfo{pages}{13485}
  (\bibinfo{year}{2016}).

\bibitem{gche2}
\bibinfo{author}{Hansen, M.~H.}, \bibinfo{author}{Jin, C.},
  \bibinfo{author}{Thygesen, K.~S.} \& \bibinfo{author}{Rossmeisl, J.}
\newblock \bibinfo{title}{$p${H} in grand canonical statistics of an
  electrochemical interface}.
\newblock \emph{\bibinfo{journal}{J. Phys. Chem. C}}
  \textbf{\bibinfo{volume}{120}}, \bibinfo{pages}{29135}
  (\bibinfo{year}{2016}).

\bibitem{agross2020}
\bibinfo{author}{Sakong, S.} \& \bibinfo{author}{Groß, A.}
\newblock \bibinfo{title}{Water structures on a {P}t(111) electrode from ab
  initio molecular dynamic simulations for a variety of electrochemical
  conditions}.
\newblock \emph{\bibinfo{journal}{Phys. Chem. Chem. Phys.}}
  \textbf{\bibinfo{volume}{22}}, \bibinfo{pages}{10431--10437}
  (\bibinfo{year}{2020}).

\bibitem{omranpoor2023}
\bibinfo{author}{Omranpoor, A.~H.} \emph{et~al.}
\newblock \bibinfo{title}{2-propanol interacting with {C}o$_3${O}$_4$(001): A
  combined v{SFS} and {AIMD} study}.
\newblock \emph{\bibinfo{journal}{J. Chem. Phys.}}
  \textbf{\bibinfo{volume}{158}}, \bibinfo{pages}{164703}
  (\bibinfo{year}{2023}).

\bibitem{cucinotta}
\bibinfo{author}{Khatib, R.}, \bibinfo{author}{Kumar, A.},
  \bibinfo{author}{Sanvito, S.}, \bibinfo{author}{Sulpizi, M.} \&
  \bibinfo{author}{Cucinotta, C.~S.}
\newblock \bibinfo{title}{The nanoscale structure of the {P}t-water double
  layer under bias revealed}.
\newblock \emph{\bibinfo{journal}{Electrochim. Acta}}
  \textbf{\bibinfo{volume}{391}} (\bibinfo{year}{2021}).

\bibitem{surendralal2018}
\bibinfo{author}{Surendralal, S.}, \bibinfo{author}{Todorova, M.},
  \bibinfo{author}{Finnis, M.~W.} \& \bibinfo{author}{Neugebauer, J.}
\newblock \bibinfo{title}{First-principles approach to model electrochemical
  reactions: Understanding the fundamental mechanisms behind {M}g corrosion}.
\newblock \emph{\bibinfo{journal}{Phys. Rev. Lett.}}
  \textbf{\bibinfo{volume}{120}}, \bibinfo{pages}{246801}
  (\bibinfo{year}{2018}).

\bibitem{shiraishi}
\bibinfo{author}{Shiraishi, K.}
\newblock \bibinfo{title}{A new slab model approach for electronic structure
  calculation of polar semiconductor surface}.
\newblock \emph{\bibinfo{journal}{J. Phys. Soc. Jpn.}}
  \textbf{\bibinfo{volume}{59}}, \bibinfo{pages}{3455--3458}
  (\bibinfo{year}{1990}).

\bibitem{deissenbeck2023}
\bibinfo{author}{Dei{\ss}enbeck, F.} \& \bibinfo{author}{Wippermann, S.}
\newblock \bibinfo{title}{Dielectric properties of nanoconfined water from ab
  initio thermopotentiostat molecular dynamics}.
\newblock \emph{\bibinfo{journal}{J. Chem. Theory Comput.}}
  \textbf{\bibinfo{volume}{19}}, \bibinfo{pages}{1035 -- 1043}
  (\bibinfo{year}{2023}).

\bibitem{dudzinski2023}
\bibinfo{author}{Dudzinski, A.~M.}, \bibinfo{author}{Diesen, E.},
  \bibinfo{author}{Heenen, H.~H.}, \bibinfo{author}{Bukas, V.~J.} \&
  \bibinfo{author}{Reuter, K.}
\newblock \bibinfo{title}{First step of the oxygen reduction reaction on
  {A}u(111): A computational study of {O}$_2$ adsorption at the electrified
  metal/water interface}.
\newblock \emph{\bibinfo{journal}{ACS Catal.}} \textbf{\bibinfo{volume}{13}},
  \bibinfo{pages}{12084 -- 12081} (\bibinfo{year}{2023}).

\bibitem{anderssonlin}
\bibinfo{author}{Andersson, L.}, \bibinfo{author}{Sprik, M.},
  \bibinfo{author}{Hutter, J.} \& \bibinfo{author}{Zhang, C.}
\newblock \bibinfo{title}{Electronic response and charge inversion at polarized
  gold electrode}.
\newblock \emph{\bibinfo{journal}{Angew. Chem. Int. Ed.}}
  \textbf{\bibinfo{volume}{64}}, \bibinfo{pages}{e202413614}
  (\bibinfo{year}{2025}).

\bibitem{freysoldt2020gdc}
\bibinfo{author}{Freysoldt, C.}, \bibinfo{author}{Mishra, A.},
  \bibinfo{author}{Ashton, M.} \& \bibinfo{author}{Neugebauer, J.}
\newblock \bibinfo{title}{Generalized dipole correction for charged surfaces in
  the repeated-slab approach}.
\newblock \emph{\bibinfo{journal}{Phys. Rev. B}}
  \textbf{\bibinfo{volume}{102}}, \bibinfo{pages}{045403}
  (\bibinfo{year}{2020}).

\bibitem{plaisance2023}
\bibinfo{author}{Islam, S. M.~R.}, \bibinfo{author}{Khezeli, F.},
  \bibinfo{author}{Ringe, S.} \& \bibinfo{author}{Plaisance, C.}
\newblock \bibinfo{title}{An implicit electrolyte model for plane wave density
  functional theory exhibiting nonlinear response and a nonlocal cavity
  definition}.
\newblock \emph{\bibinfo{journal}{J. Chem. Phys.}}
  \textbf{\bibinfo{volume}{159}}, \bibinfo{pages}{234117}
  (\bibinfo{year}{2023}).

\bibitem{environ1}
\bibinfo{author}{Andreussi, O.}, \bibinfo{author}{Dabo, I.} \&
  \bibinfo{author}{Marzari, N.}
\newblock \bibinfo{title}{Revised self-consistent continuum solvation in
  electronic-structure calculations}.
\newblock \emph{\bibinfo{journal}{J. Chem. Phys.}}
  \textbf{\bibinfo{volume}{136}}, \bibinfo{pages}{064102}
  (\bibinfo{year}{2012}).

\bibitem{environ2}
\bibinfo{author}{Giannozzi, P.} \emph{et~al.}
\newblock \bibinfo{title}{Advanced capabilities for materials modelling with
  {Q}uantum {ESPRESSO}}.
\newblock \emph{\bibinfo{journal}{J. Phys.: Cond. Matter}}
  \textbf{\bibinfo{volume}{29}}, \bibinfo{pages}{465901}
  (\bibinfo{year}{2017}).

\bibitem{otani2017}
\bibinfo{author}{Nishihara, S.} \& \bibinfo{author}{Otani, M.}
\newblock \bibinfo{title}{Hybrid solvation models for bulk, interface, and
  membrane: Reference interaction site methods coupled with density functional
  theory}.
\newblock \emph{\bibinfo{journal}{Phys. Rev. B}} \textbf{\bibinfo{volume}{96}},
  \bibinfo{pages}{115429} (\bibinfo{year}{2017}).

\bibitem{otani2022}
\bibinfo{author}{Hagiwara, S.}, \bibinfo{author}{Nishihara, S.},
  \bibinfo{author}{Kuroda, F.} \& \bibinfo{author}{Otani, M.}
\newblock \bibinfo{title}{Development of a dielectrically consistent reference
  interaction site model combined with the density functional theory for
  electrochemical interface simulations}.
\newblock \emph{\bibinfo{journal}{Phys. Rev. Mater.}}
  \textbf{\bibinfo{volume}{6}}, \bibinfo{pages}{093802}
  (\bibinfo{year}{20222}).

\bibitem{arias2017}
\bibinfo{author}{Sundararaman, R.}, \bibinfo{author}{Goddard, W.~A., III} \&
  \bibinfo{author}{Arias, T.~A.}
\newblock \bibinfo{title}{Grand canonical electronic density-functional theory:
  Algorithms and applications to electrochemistry}.
\newblock \emph{\bibinfo{journal}{J. Chem. Phys.}}
  \textbf{\bibinfo{volume}{146}}, \bibinfo{pages}{114104}
  (\bibinfo{year}{2017}).

\bibitem{cucinotta2024}
\bibinfo{author}{Buraschi, M.}, \bibinfo{author}{Horsfield, A.~P.} \&
  \bibinfo{author}{Cucinotta, C.~S.}
\newblock \bibinfo{title}{Revealing interface polarization effects on the
  electrical doublelayer with efficient open boundary simulations under
  potentialcontrol}.
\newblock \emph{\bibinfo{journal}{J. Phys. Chem. Lett.}}
  \textbf{\bibinfo{volume}{15}}, \bibinfo{pages}{4872--4879}
  (\bibinfo{year}{2024}).

\bibitem{otani2012PRL}
\bibinfo{author}{Bonnet, N.}, \bibinfo{author}{Morishita, T.},
  \bibinfo{author}{Sugino, O.} \& \bibinfo{author}{Otani, M.}
\newblock \bibinfo{title}{First-principles molecular dynamics at a constant
  electrode potential}.
\newblock \emph{\bibinfo{journal}{Phys. Rev. Lett.}}
  \textbf{\bibinfo{volume}{109}}, \bibinfo{pages}{266101}
  (\bibinfo{year}{2012}).

\bibitem{bouzid2017}
\bibinfo{author}{Bouzid, A.} \& \bibinfo{author}{Pasquarello, A.}
\newblock \bibinfo{title}{Redox levels through constant {F}ermi-level ab initio
  molecular dynamics}.
\newblock \emph{\bibinfo{journal}{J. Chem. Theory Comput.}}
  \textbf{\bibinfo{volume}{13}}, \bibinfo{pages}{1769--1777}
  (\bibinfo{year}{2017}).

\bibitem{deissenbeck2021}
\bibinfo{author}{Dei{\ss}enbeck, F.}, \bibinfo{author}{Freysoldt, C.},
  \bibinfo{author}{Todorova, M.}, \bibinfo{author}{Neugebauer, J.} \&
  \bibinfo{author}{Wippermann, S.}
\newblock \bibinfo{title}{Dielectric properties of nanoconfined water: A
  canonical thermopotentiostat approach}.
\newblock \emph{\bibinfo{journal}{Phys. Rev. Lett.}}
  \textbf{\bibinfo{volume}{126}}, \bibinfo{pages}{136803}
  (\bibinfo{year}{2021}).

\bibitem{sprikPRB}
\bibinfo{author}{Zhang, C.} \& \bibinfo{author}{Sprik, M.}
\newblock \bibinfo{title}{Computing the dielectric constant of liquid water at
  constant dielectric displacement}.
\newblock \emph{\bibinfo{journal}{Phys. Rev. B}} \textbf{\bibinfo{volume}{93}},
  \bibinfo{pages}{144201} (\bibinfo{year}{2016}).

\bibitem{yoo2021}
\bibinfo{author}{Yoo, S.-H.} \emph{et~al.}
\newblock \bibinfo{title}{Finite-size correction for slab supercell
  calculations of materials with spontaneous polarization}.
\newblock \emph{\bibinfo{journal}{npj Comput. Mater.}}
  \textbf{\bibinfo{volume}{7}}, \bibinfo{pages}{58} (\bibinfo{year}{2021}).

\bibitem{melander_dyn}
\bibinfo{author}{Melander, M.~M.}
\newblock \bibinfo{title}{Frozen or dynamic? — {A}n atomistic simulation
  perspective on the timescales of electrochemical reactions}.
\newblock \emph{\bibinfo{journal}{Electrochim. Acta}}
  \textbf{\bibinfo{volume}{446}}, \bibinfo{pages}{142095}
  (\bibinfo{year}{2023}).

\bibitem{marcus2009}
\bibinfo{author}{Maurice, V.}, \bibinfo{author}{Klein, L.~H.},
  \bibinfo{author}{Strehblow, H.-H.} \& \bibinfo{author}{Marcus, P.}
\newblock \bibinfo{title}{In situ {STM} study of the surface structure,
  dissolution, and early stages of electrochemical oxidation of the {A}g(111)
  electrode}.
\newblock \emph{\bibinfo{journal}{J. Phys. Chem. C}}
  \textbf{\bibinfo{volume}{111}}, \bibinfo{pages}{16351--16361}
  (\bibinfo{year}{2007}).

\bibitem{Huang2020Mg}
\bibinfo{author}{Huang, J.}, \bibinfo{author}{Song, G.-L.},
  \bibinfo{author}{Atrens, A.} \& \bibinfo{author}{Dargusch, M.}
\newblock \bibinfo{title}{What activates the {M}g surface -- a comparison of
  {M}g dissolution mechanisms}.
\newblock \emph{\bibinfo{journal}{J. Mater. Sci. Technol.}}
  \textbf{\bibinfo{volume}{57}}, \bibinfo{pages}{204--220}
  (\bibinfo{year}{2020}).

\bibitem{Yuwono2019}
\bibinfo{author}{Yuwono, J.~A.} \emph{et~al.}
\newblock \bibinfo{title}{Aqueous electrochemistry of the magnesium surface:
  Thermodynamic and kinetic profiles}.
\newblock \emph{\bibinfo{journal}{Corr. Sci.}} \textbf{\bibinfo{volume}{147}},
  \bibinfo{pages}{53--68} (\bibinfo{year}{2020}).

\bibitem{rossmeisl}
\bibinfo{author}{Rossmeisl, J.}, \bibinfo{author}{Skulason, E.},
  \bibinfo{author}{Bj\"orketun, M.~E.}, \bibinfo{author}{Tripkovic, V.} \&
  \bibinfo{author}{Norskov, J.~K.}
\newblock \bibinfo{title}{Modeling the electrified solid–liquid interface}.
\newblock \emph{\bibinfo{journal}{Chem. Phys. Lett.}}
  \textbf{\bibinfo{volume}{466}}, \bibinfo{pages}{68} (\bibinfo{year}{2008}).

\bibitem{janik2009}
\bibinfo{author}{Janik, M.~J.}, \bibinfo{author}{Taylor, C.~D.} \&
  \bibinfo{author}{Neurock, M.}
\newblock \bibinfo{title}{First-principles analysis of the initial
  electroreduction steps of oxygen over {P}t(111)}.
\newblock \emph{\bibinfo{journal}{J. Electrochem. Soc.}}
  \textbf{\bibinfo{volume}{156}}, \bibinfo{pages}{B126--B135}
  (\bibinfo{year}{2008}).

\bibitem{laasonen}
\bibinfo{author}{Kronberg, R.} \& \bibinfo{author}{Laasonen, K.}
\newblock \bibinfo{title}{Coupling surface coverage and electrostatic effects
  on the interfacial adlayer–water structure of hydrogenated single-crystal
  platinum electrodes}.
\newblock \emph{\bibinfo{journal}{J. Phys. Chem. C}}
  \textbf{\bibinfo{volume}{124}}, \bibinfo{pages}{13706--13714}
  (\bibinfo{year}{2020}).

\bibitem{santos2022}
\bibinfo{author}{Santos, E.} \& \bibinfo{author}{Schmickler, W.}
\newblock \bibinfo{title}{Models of electron transfer at different electrode
  materials}.
\newblock \emph{\bibinfo{journal}{Chem. Rev.}} \textbf{\bibinfo{volume}{122}},
  \bibinfo{pages}{10581-- 10598} (\bibinfo{year}{2022}).

\bibitem{flucts}
\bibinfo{author}{Wippermann, S.}, \bibinfo{author}{Todorova, M.} \&
  \bibinfo{author}{Neugebauer, J.}
\newblock \bibinfo{title}{(in preparation)} .

\bibitem{surendralal2021}
\bibinfo{author}{Surendralal, S.}, \bibinfo{author}{Todorova, M.} \&
  \bibinfo{author}{Neugebauer, J.}
\newblock \bibinfo{title}{Impact of water coadsorption on the electrode
  potential of {H}-{P}t(111)-liquid water interfaces}.
\newblock \emph{\bibinfo{journal}{Phys. Rev. Lett.}}
  \textbf{\bibinfo{volume}{126}}, \bibinfo{pages}{166802}
  (\bibinfo{year}{2021}).

\bibitem{pzc_jle}
\bibinfo{author}{Le, J.}, \bibinfo{author}{Ianuzzi, M.},
  \bibinfo{author}{Cuesta, A.} \& \bibinfo{author}{Cheng, J.}
\newblock \bibinfo{title}{Determining potentials of zero charge of metal
  electrodes versus the standard hydrogen electrode from
  density-functional-theory-based molecular dynamics}.
\newblock \emph{\bibinfo{journal}{Phys. Rev. Lett.}}
  \textbf{\bibinfo{volume}{119}}, \bibinfo{pages}{016801}
  (\bibinfo{year}{2017}).

\bibitem{sundararaman2022}
\bibinfo{author}{Shandilya, A.}, \bibinfo{author}{Schwarz, K.} \&
  \bibinfo{author}{Sundararaman, R.}
\newblock \bibinfo{title}{Interfacial water asymmetry at ideal electrochemical
  interfaces}.
\newblock \emph{\bibinfo{journal}{J. Chem. Phys.}}
  \textbf{\bibinfo{volume}{156}}, \bibinfo{pages}{014705}
  (\bibinfo{year}{2022}).

\bibitem{PhD_sudarsan}
\bibinfo{author}{Surendralal, S.}
\newblock \emph{\bibinfo{title}{Development of an ab initio computational
  potentiostat and its application to the study of {M}g corrosion}}.
\newblock Ph.D. thesis, \bibinfo{school}{Ruhr-Universit\"at Bochum}
  (\bibinfo{year}{2020}).
\newblock
  \urlprefix\url{https://hss-opus.ub.ruhr-uni-bochum.de/opus4/frontdoor/index/index/year/2020/docId/7010}.

\bibitem{spohr}
\bibinfo{author}{Omranpoor, A.}, \bibinfo{author}{Kox, T.},
  \bibinfo{author}{Spohr, E.} \& \bibinfo{author}{Kenmoe, S.}
\newblock \bibinfo{title}{Influence of temperature, surface composition and
  electrochemical environment on 2-propanol decomposition at the
  {C}o$_3${O}$_4$(001)/{H}$_2${O} interface}.
\newblock \emph{\bibinfo{journal}{Appl. Surf. Sci. Adv.}}
  \textbf{\bibinfo{volume}{12}}, \bibinfo{pages}{100319}
  (\bibinfo{year}{2022}).

\bibitem{umbrella}
\bibinfo{author}{Torrie, G.} \& \bibinfo{author}{Valleau, J.}
\newblock \bibinfo{title}{Nonphysical sampling distributions in monte carlo
  free-energy estimation: Umbrella sampling}.
\newblock \emph{\bibinfo{journal}{J. Comput. Phys.}}
  \textbf{\bibinfo{volume}{23}}, \bibinfo{pages}{187--199}
  (\bibinfo{year}{1977}).

\bibitem{metadyn1}
\bibinfo{author}{Laio, A.} \& \bibinfo{author}{Parrinello, M.}
\newblock \bibinfo{title}{Escaping free-energy minima}.
\newblock \emph{\bibinfo{journal}{Proc. Nat. Acad. Sci.}}
  \textbf{\bibinfo{volume}{99}}, \bibinfo{pages}{12562--12566}
  (\bibinfo{year}{2002}).

\bibitem{metadyn2}
\bibinfo{author}{Tiwary, P.} \& \bibinfo{author}{Parrinello, M.}
\newblock \bibinfo{title}{From metadynamics to dynamics}.
\newblock \emph{\bibinfo{journal}{Phys. Rev. Lett.}}
  \textbf{\bibinfo{volume}{111}}, \bibinfo{pages}{230602}
  (\bibinfo{year}{2013}).

\bibitem{Allen2006}
\bibinfo{author}{Allen, R.~J.}, \bibinfo{author}{Frenkel, D.} \&
  \bibinfo{author}{ten Wolde, P.~R.}
\newblock \bibinfo{title}{Forward flux sampling-type schemes for simulating
  rare events: Efficiency analysis}.
\newblock \emph{\bibinfo{journal}{J. Chem. Phys.}}
  \textbf{\bibinfo{volume}{124}}, \bibinfo{pages}{194111}
  (\bibinfo{year}{2006}).

\bibitem{Allen2009}
\bibinfo{author}{Allen, R.~J.}, \bibinfo{author}{Valeriani, C.} \&
  \bibinfo{author}{Rein~ten Wolde, P.}
\newblock \bibinfo{title}{Forward flux sampling for rare event simulations}.
\newblock \emph{\bibinfo{journal}{J. Phys.: Cond. Matt.}}
  \textbf{\bibinfo{volume}{21}}, \bibinfo{pages}{463102}
  (\bibinfo{year}{2009}).

\bibitem{bluemoon}
\bibinfo{author}{Ciccotti, G.}, \bibinfo{author}{Kapral, R.} \&
  \bibinfo{author}{Vanden-Eijnden, E.}
\newblock \bibinfo{title}{Blue moon sampling, vectorial reaction coordinates,
  and unbiased constrained dynamics}.
\newblock \emph{\bibinfo{journal}{Chem. Phys. Chem.}}
  \textbf{\bibinfo{volume}{6}}, \bibinfo{pages}{1809--1814}
  (\bibinfo{year}{2005}).

\bibitem{csvr}
\bibinfo{author}{Bussi, G.}, \bibinfo{author}{Donadio, D.} \&
  \bibinfo{author}{Parrinello, M.}
\newblock \bibinfo{title}{{Canonical sampling through velocity rescaling}}.
\newblock \emph{\bibinfo{journal}{J. Chem. Phys.}}
  \textbf{\bibinfo{volume}{126}}, \bibinfo{pages}{014101}
  (\bibinfo{year}{2007}).

\bibitem{landau}
\bibinfo{author}{Landau, L.~D.} \emph{et~al.}
\newblock \emph{\bibinfo{title}{Electrodynamics of continuous media}}
  Vol.~\bibinfo{volume}{8} (\bibinfo{publisher}{Elsevier},
  \bibinfo{year}{2013}).

\bibitem{henkel16}
\bibinfo{author}{Kumph, M.}, \bibinfo{author}{Henkel, C.},
  \bibinfo{author}{Rabl, P.}, \bibinfo{author}{Brownnutt, M.} \&
  \bibinfo{author}{Blatt, R.}
\newblock \bibinfo{title}{Electric-field noise above a thin dielectric layer on
  metal electrodes}.
\newblock \emph{\bibinfo{journal}{New J. Phys.}} \textbf{\bibinfo{volume}{18}},
  \bibinfo{pages}{023020} (\bibinfo{year}{2016}).

\bibitem{gardiner}
\bibinfo{author}{Gardiner, C.}
\newblock \emph{\bibinfo{title}{Stochastic Methods}}
  (\bibinfo{publisher}{Springer}, \bibinfo{address}{Berlin},
  \bibinfo{year}{2009}).

\bibitem{dellago2023}
\bibinfo{author}{Omranpour, A.}, \bibinfo{author}{De~Hijes, P.~M.},
  \bibinfo{author}{Behler, J.} \& \bibinfo{author}{Dellago, C.}
\newblock \bibinfo{title}{Perspective: Atomistic simulations of water and
  aqueous systems with machine learning potentials}.
\newblock \emph{\bibinfo{journal}{J. Chem. Phys.}}
  \textbf{\bibinfo{volume}{160}}, \bibinfo{pages}{170901}
  (\bibinfo{year}{2024}).

\bibitem{Jung2023}
\bibinfo{author}{Jung, J.~H.}, \bibinfo{author}{Srinivasan, P.},
  \bibinfo{author}{Forslund, A.} \& \bibinfo{author}{Grabowski, B.}
\newblock \bibinfo{title}{High-accuracy thermodynamic properties to the melting
  point from ab initio calculations aided by machine-learning potentials}.
\newblock \emph{\bibinfo{journal}{npj Comput. Mater.}}
  \textbf{\bibinfo{volume}{9}}, \bibinfo{pages}{3} (\bibinfo{year}{2023}).

\end{thebibliography}

\end{document}